\documentclass[11pt]{article}

\usepackage[margin=1in]{geometry}
\usepackage[dvipsnames]{xcolor}
\usepackage[numbers,sort&compress]{natbib}
\usepackage{authblk}
\usepackage{booktabs}
\usepackage[section]{placeins}  
\usepackage{libertinus} 
\usepackage{algorithm}
\usepackage[noend]{algpseudocode}
\usepackage[hidelinks]{hyperref}
\usepackage{wrapfig}  
\usepackage{bbm}
\usepackage{orcidlink}

\usepackage{etoolbox}
\newtoggle{submission}
\togglefalse{submission} 

\usepackage{amssymb,amsmath,amsfonts,amsthm}
\usepackage{graphicx}
\usepackage{subcaption}
\usepackage{enumitem}
\usepackage{parskip}
\usepackage{ifthen}
\usepackage{ascii}
\usepackage{newunicodechar}
\usepackage{bbding}
\usepackage{float}
\usepackage{array}  
\usepackage{tikz}
\usepackage{bm}
\usepackage{bbm}
\usepackage{lineno}
\usepackage{empheq}
\usetikzlibrary{arrows.meta,backgrounds,fit,positioning,shapes.geometric}

\usepackage[most]{tcolorbox}
\usepackage{tikz}
\usetikzlibrary{arrows.meta,backgrounds,fit,positioning,shapes.geometric}
\definecolor{sidebarblue}{HTML}{315B73}
\definecolor{sidebarbackground}{HTML}{F2F6F8}
\newtcolorbox{sidebar}[1][]{
    enhanced,
    breakable,
    colback=sidebarbackground,
    colframe=sidebarbackground,
    boxrule=0pt,
    borderline west={2.5pt}{0pt}{sidebarblue},
    sharp corners,
    left=7pt,
    right=7pt,
    top=6pt,
    bottom=6pt,
    before skip=8pt,
    after skip=8pt,
    fonttitle=\bfseries\small,
    coltitle=sidebarblue,
    title={#1}
}

\usepackage[most]{tcolorbox}
\usepackage{amsmath}
\usepackage{cleveref}          

\definecolor{latentink}{HTML}{8967AF}   
\definecolor{moverink}{HTML}{2F7D52}    
\definecolor{stayerink}{HTML}{5B6065}   
\definecolor{paperink}{HTML}{1A1A1A}    
\definecolor{asidebg}{HTML}{F4F4F4}     

\newcommand{\latent}[1]{#1}
\newcommand{\mover}[1]{#1}
\newcommand{\stayer}[1]{#1}

\newcounter{sidebarbox}
\newcounter{savedeqno}
\providecommand{\sidebarname}{Sidebar}
\crefname{sidebarbox}{Sidebar}{Sidebars}
\Crefname{sidebarbox}{Sidebar}{Sidebars}

\theoremstyle{remark}

\DeclareMathOperator*{\argmax}{arg\,max}

\newcommand{\brackets}[1]{\left[#1\right]}

\newcommand{\abs}[1]{\left\lvert #1 \right\rvert}

\newcommand{\E}[2][]{
    \ifthenelse{\equal {#1} {}}
    {
        \ifthenelse{\equal {#2} {}}
        {\mathbb{E}}
        {\mathbb{E}\brackets{#2}}
    }
    {
        \ifthenelse{\equal {#2} {}}
        {\mathbb{E}_{#1}}
        {\mathbb{E}_{#1}\brackets{#2}}
    }
}

\newcommand{\probsym}{\mathbb{P}}
\newcommand{\prob}[1]{
    \ifthenelse{\equal {#1} {}}
        {\probsym}
        {\probsym\brackets{#1}}
}

\newcommand{\param}{\boldsymbol{\phi}}

\newcommand{\kernel}{k}

\newcommand{\nmSup}[1]{\ifx\relax#1\relax\else^{(#1)}\fi}  
\newcommand{\nmSub}[1]{\ifx\relax#1\relax\else_{#1}\fi}    

\newcommand{\glyphEvent}{e}               
\newcommand{\glyphSource}{s}              
\newcommand{\glyphSourceSet}{\mathcal{S}} 
\newcommand{\glyphEMiter}{\tau}           
\newcommand{\glyphOpinion}{x}             
\newcommand{\glyphTarget}{i}              

\newcommand{\glyphPull}{d}        
\newcommand{\glyphRadius}{\lvert \pull \rvert}      

\newcommand{\glyphJointResp}{\bar{z}} 
\newcommand{\glyphDev}{\xi}           
\newcommand{\glyphUpd}{\Delta}        
\newcommand{\glyphUpdObs}{\delta}     

\newcommand{\phiz}{\phi_{0}}     
\newcommand{\phio}{\phi_{1}}     
\newcommand{\glyphWidth}{\phiz}  
\newcommand{\glyphSteep}{\phio}  
\newcommand{\scalarparam}{\phi}  

\newcommand{\ev}{\glyphEvent}                              
\newcommand{\src}{\glyphSource}                            
\newcommand{\es}{\glyphEvent,\glyphSource}                 
\newcommand{\srcset}[1][\glyphEvent]{\glyphSourceSet_{#1}} 
\newcommand{\emiter}{\glyphEMiter}                         

\newcommand{\opre}[1][\glyphEvent]{\glyphOpinion_{#1}^{-}}   
\newcommand{\opost}[1][\glyphEvent]{\glyphOpinion_{#1}^{+}}  

\newcommand{\pull}[1][\glyphEvent]{\glyphPull_{#1}}     
\newcommand{\radius}[1][\glyphEvent]{\left\lvert \pull[#1] \right\rvert} 
\newcommand{\gainat}[1]{\kernel_{\param}(#1)}           
\newcommand{\gain}[1][\glyphEvent]{\gainat{\radius[#1]}} 
\newcommand{\jointresp}[1][\es]{\glyphJointResp_{#1}}   

\newcommand{\updrv}[1][\glyphEvent]{\glyphUpd_{#1}}   
\newcommand{\upd}[1][\glyphEvent]{\glyphUpdObs_{#1}}  
\newcommand{\movind}[1][\glyphEvent]{z_{#1}}          
\newcommand{\dev}[1][\glyphSource]{\glyphDev_{#1}}    

\definecolor{mygreen}{rgb}{0.1, 0.5, 0.1}

\newcommand{\vkg}{\texttt{VK-gauging}}      
\newcommand{\vkc}{\texttt{VK-counting}}     
\newcommand{\vk}{\texttt{VK}}               
\newcommand{\kozitsin}{\texttt{VKontakte}}  
\newcommand{\senate}{\texttt{SENATE}}       
\newcommand{\house}{\texttt{HOUSE}}         
\newcommand{\cmv}{\texttt{CMV}}             
\newcommand{\spinos}{\texttt{SPINOS}}       
\newcommand{\markets}{\texttt{MARKETS}}     

\newcommand{\adams}{\texttt{ADAMS}}                  
\newcommand{\adamsLo}{\texttt{ADAMS-low}}            
\newcommand{\adamsMid}{\texttt{ADAMS-medium}}         
\newcommand{\adamsHi}{\texttt{ADAMS-high}}           
\newcommand{\becker}{\texttt{BECKER}}                
\newcommand{\takacs}{\texttt{TAKACS}}                
\newcommand{\takacsOne}{\texttt{TAKACS-S1}}          
\newcommand{\takacsCtl}{\texttt{TAKACS-S2-control}}  
\newcommand{\takacsDis}{\texttt{TAKACS-S2-disliking}} 
\newcommand{\takacsInt}{\texttt{TAKACS-interior}}    

\newcommand{\rff}{\texttt{RFF}}
\newcommand{\sar}{\texttt{SAR}}

\usepackage{bm}

\title{Learning Social Influence with Unobserved Attention}

\author[1]{William Thompson\orcidlink{0000-0002-1893-240X}\thanks{Corresponding author: wthomps3@uvm.edu}}
\author[2]{Moyi Tian\orcidlink{0000-0001-8166-7410}}
\author[3]{Linh Huynh\orcidlink{0000-0003-0540-6811}}
\author[4]{Alexander Wiedemann\orcidlink{0000-0001-5156-1084}}
\author[5]{Vicki Modisette\orcidlink{0009-0006-3254-3175}}
\author[6]{Philip S. Chodrow\orcidlink{0000-0001-5667-490X}}
\author[7]{Heather Z. Brooks\orcidlink{0000-0003-4274-107X}}

\affil[1]{Complex Systems Institute, University of Vermont}
\affil[2]{Department of Applied Mathematics, University of Colorado Boulder}
\affil[3]{Honors College Applied Mathematics, Texas Tech University}
\affil[4]{Department of Mathematics and Computational Data Science, Hamline University}
\affil[5]{Department of Mathematics and Statistics, Eastern Kentucky University}
\affil[6]{Department of Computer Science, Middlebury College}
\affil[7]{Department of Mathematics, Harvey Mudd College}

\date{}

\begin{document}

\maketitle

\begin{abstract}
    Social influence, the tendency of our social environment to sway our beliefs and behaviors, is often cited as a driver of population-level trends in opinions or beliefs.
Many quantitative theories of social influence assume that an agent interacts with its entire social environment.
This assumption is often unrealistic. For example, users of online platforms typically see posts from only a small subset of the accounts they follow at any given time.
We introduce a simple probabilistic agent-based model of social influence that explicitly models unobserved attention mechanisms. The first of these latent mechanisms is \emph{selection}: each agent interacts with only one other agent at a time from its social environment. The second is \emph{retention}: the agent chooses whether or not to ignore the opinion of the other.
Our model admits an efficient maximum-likelihood algorithm which allows us to fit it to opinion traces even when these mechanisms are unobserved.
On synthetic data, omitting either latent mechanism reduces the overall strength of social influence governed by the inferred interaction kernel, which hides the evidence for the underlying nonlinearity of the social influence.
We study a selection of previously analyzed empirical datasets and curate several new ones.
We show that models that include the latent mechanisms are often more predictive on these datasets; indeed, they typically improve prediction far more than the choice of interaction kernel does.
Modeling latent attention increases held-out predictive likelihood on 14 of these 15 datasets.
Each mechanism changes the size of the inferred kernel in a predictable direction, while its effect on the kernel's shape depends on the data.

\end{abstract}

The idea that people influence each other's beliefs, opinions and practices is well-established in social psychology \cite{sherif1936}.
The mechanisms of this influence are still an active area of study, and so are formal mathematical models of it.

Mathematical models of opinion dynamics \cite{hegselmann2002,Castellano2009b,Flache2017,Noorazar2020,peralta2025opinion} are an important class of interacting particle systems \cite{liggett1985interacting,fengLearningInteractionVariables2022} for studying social influence.
Models of opinion dynamics give deterministic or stochastic rules by which an agent updates their opinion in response to the opinions of other agents in their environment. In some cases, these agents are placed on a social network which limits who can influence whom; in other cases all agents may interact with each other.

These models often assume that an agent is influenced by \emph{every} other agent in their local environment at any given point in time. Several psychological studies make the same assumption inferring social influence as a function of the average opinion of all available neighbors~\cite{VandeKerckhove2016,Kozitsin2021,Becker2017,Becker2019}.
It is also common in models of opinion dynamics to assume that agents must be influenced by every opinion they are exposed to, even though controlled experiments often find that the most common response to exposure is to maintain one's current opinion \cite{adams2022person,Hovland1957,Takacs2016,moussaid2013social,Kozitsin2021}.
These two assumptions are natural in the context of data analysis because they make the data we can observe sufficient to reconstruct social influence: We \emph{can} observe the opinions of agents over time and the social network connecting them. We \emph{cannot} easily observe who an agent attended to or if that exposure changed their opinion.

These assumptions are unrealistic in at least some settings.
Consider a social media user who scrolls through a platform for a few minutes.
The underlying social network (e.g.,\ the follower graph) may include hundreds of neighbors. The user cannot be influenced by all of them at once, because only a small fraction of their posts appear in the user's feed.
The user may also not be open to changing their opinion on a given topic at all during that session.
What the user \emph{pays attention to}, whether through algorithmic curation, user choice, or some other mechanism, shapes their opinion update. Empirical data typically do not record it.

\begin{figure*}[t]
    \centering
    \includegraphics[width=\textwidth]{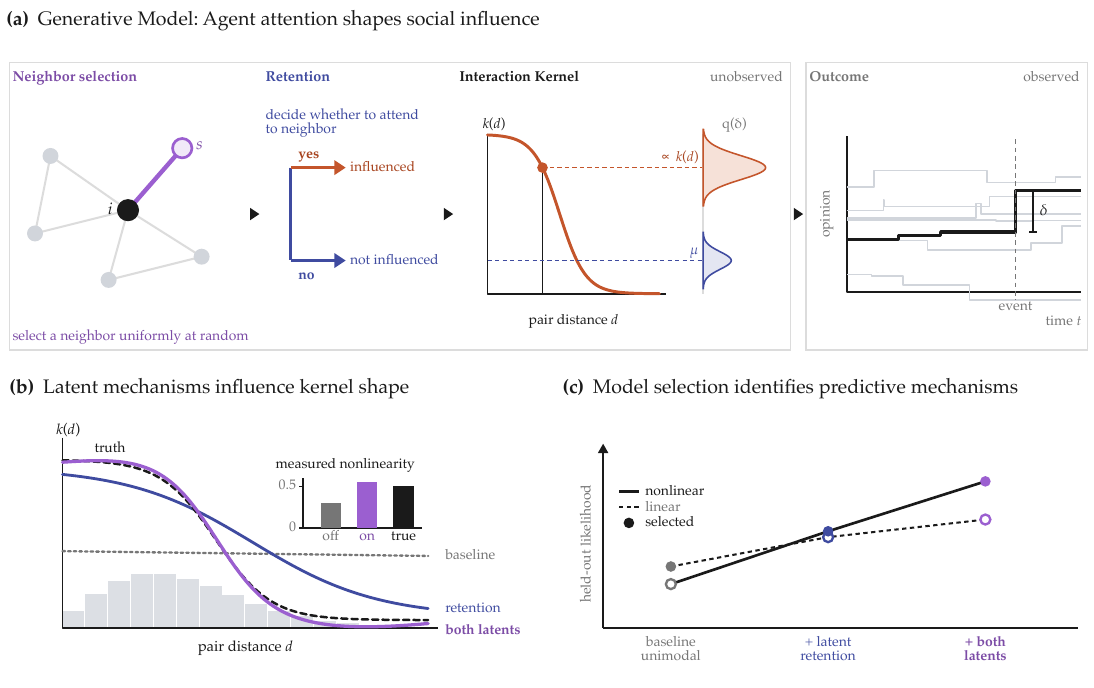}
    \caption{
    A schematic illustrating our model, model selection pipeline, and stylized results. \textbf{(a) } Generative model. Each agent holds a scalar-valued opinion $\glyphOpinion$ in an information environment that may contain several candidate sources.
    First, the agent chooses a single source $\src$, associated with an event in the information environment, and decides whether to attend to it. If the agent attends to the source, the expected change in its opinion is proportional to the influence specified by an interaction kernel, $\kernel (\cdot)$, which depends on the distance $d$ between the agent and the source. Otherwise, the expected change is given by a drift term $\mu$, capturing opinion change unrelated to the source. The agent then makes a noisy opinion update by $\glyphUpdObs$, drawn from a distribution $q(\cdot)$.
    We observe each agent's sequence of updates, but not the sources it selects or its decisions to attend to them.
    \textbf{(b)} We fit models with and without the latent mechanisms to synthetic and empirical data, and select the model with the highest held-out likelihood. On empirical data, we select between linear (dashed) and nonlinear (solid) models by held-out log likelihood. Without the latent mechanisms, linear models are often selected over, or tied with, nonlinear models. Nonlinear models are selected on many datasets once the mechanisms are added.
    \textbf{(c)} On synthetic data where the latent mechanisms are truly present, fits that omit them reduce the overall strength of the interaction kernel and obscure evidence of nonlinearity. Adding the mechanisms restores the kernel's overall strength and its curvature (when in the support of the data).}
    \label{fig:kernel_comparison_dyadic}
\end{figure*}

We incorporate these unobserved attention mechanisms into mathematical models of social influence.
Our approach starts from a probabilistic generative agent-based model (PGABM).
A probabilistic generative model can simulate synthetic data and can compute the likelihood of observed data \citep{Monti2023,Monti2020,lenti2025variationalinference}.
It therefore supports likelihood-based parameter inference and model selection, which are often preferable to simulation-based methods \citep{Lenti2024} or curve fitting in terms of accuracy and computational efficiency. 
Our PGABM explicitly models two latent attention mechanisms. \emph{Source selection} determines which source in the information environment the agent attends to. \emph{Retention} determines whether an agent is influenced at all by a given exposure. On a network, the candidate sources are the agent's neighbors, so source selection is neighbor selection; we use the two terms interchangeably.
Most datasets do not observe these mechanisms. We show that expectation-maximization, a standard method for latent-variable models, fits our PGABM to empirical data efficiently. 
We illustrate our generative model and the model selection pipeline in \Cref{fig:kernel_comparison_dyadic}.

We introduce our model in \Cref{sec:model}, focusing on how we can combine classical opinion dynamics models with the latent attention mechanisms.
In \Cref{sec:inference} we describe our inference and model selection pipeline.
In \Cref{sec:results} we apply these methods to synthetic and empirical data.

On synthetic data, omitting a latent mechanism that is truly present in the data-generating process biases the inferred kernel.
It reduces the overall strength of the kernel, which in turn reduces identifiability.
On empirical data, we find strong evidence for the latent mechanisms across a wide range of datasets. We apply our pipeline to several previously analyzed controlled experiments as well as new datasets curated for this study.
Held-out predictive likelihoods show that \emph{modeling latent attention mechanisms improves the predictive capability of models}.
We then track the overall strength and curvature of the inferred kernels. Each mechanism changes the strength in a predictable direction, while its effect on the curvature depends on the data. Together, the mechanisms change the verdict on whether the interaction kernel is linear or nonlinear on more than half of the datasets.
We close with a discussion and suggestions for future work in \Cref{sec:discussion}.

\section{Probabilistic Generative Agent-Based Model (PGABM)}
\label{sec:model}

Our model starts from the idea that agents do not attend equally to every signal available to them.
For example, an agent may have 100 friends on a social media platform but see posts from only a few of them on a given day. An agent may also be exposed to a source that has no systematic influence on their current opinion.

We consider an \emph{information environment} in which a population $V$ of $N = |V|$ individuals live and can be influenced. In a controlled lab study, the environment may be a single stimulus shown to participants. In a social network, it may consist of messages, posts on a timeline, or opinions communicated by other individuals. In that case, the individuals are connected by a social network $G=(V, E)$, where $E$ is the set of undirected or directed edges. The stimulus to agent $i \in V$ may then come from its neighborhood $\partial_i=\{\,j\in V \mid (i,j)\in E\,\}$. Let $\mathcal{E}$ be a discrete, ordered set of potential opinion-update events. Let the information environment associated with each event $e\in\mathcal{E}$ be $\srcset$, and at such an event $\glyphEvent$, a single focal target $i_e \in V$ may change its opinion in response to one or more potential sources $s_e \in \srcset$. The subscript $e$ marks association with event $e$.

The index $e$ orders events but not necessarily time. Asynchronous records already arrive as a sequence. Synchronous rounds are unrolled into one event per agent per round. Each event's candidate sources are the stimuli that agent saw in the round, so $|\mathcal{E}| = NT$ over $T$ rounds.

Each target has a quantifiable opinion before and after each event. It may be a numerical estimate, a survey response on a Likert scale, or a quantity inferred from behavior, such as bets in a prediction market or the text of Reddit posts. For the focal individual $i_e \in V$, we write $x_{i_e}^-$ and $x_{i_e}^+$ for their opinion immediately before and after event $e$. For notational clarity we abbreviate the focal target's opinions by
\[
\opre \equiv x_{i_e}^-, \qquad \opost \equiv x_{i_e}^+.
\]
The observed opinion update is then
\[ \upd = \opost-\opre. \]

Two parts of each update are typically unobserved: which source $s_e$ the target attends to, and whether the target responds to it at all. We treat both as latent variables. The subsections below describe how we model attention and social influence in the opinion update $\delta_e$.

\subsection{Selection of information source} \label{model:selection}

When a target's information environment contains several potential sources, we may not know which one it attends to. This is common in real-world social networks, where a user's information environment may hold dozens or hundreds of others. Empirical studies of networked social influence often average over the neighborhood \cite{VandeKerckhove2016,Kozitsin2021,Becker2017,Becker2019}, but averaging over all potential sources can hide the true mechanism. For example, a target may move toward the source it attends to while moving away from the average opinion of all sources. Averaging then makes positive influence look negative. Both \citet{Mas2013b} and \citet{Kozitsin2021} identify this as a possible source of spurious negative influence. Averaging over all sources is also psychologically implausible, given limited short-term memory \cite{Cowan2001}, selective attention \cite{Weng2012}, influence concentrated among close friends \cite{bond2012} and diminishing returns to exposure \cite{Latane1981}.

We therefore separate an individual's potential information environment from the source it actually attends to. At event $e$, the target $i_e$ attends to a single source $s_e$ from its information environment. For simplicity, we assume $s_e$ is drawn uniformly at random from the target's available sources, and we treat its identity as latent when it is not observed. Given this source, the target updates as a function of the discrepancy $\pull = x_{s_e}^- - x_{i_e}^-$. We leave more general attention rules for future work.

\subsection{Retention}
\label{model:retention}

Even after selecting a source, a target may not respond to it. Several studies report retention (no change in opinion after a stimulus) as the most common response \cite{moussaid2013social,adams2022person,Takacs2016,Chacoma2015}. Later analyses treat retention and social influence as separate deterministic modes, and use hard filters to sort transitions into one or the other. Chacoma et al.~\citep{Chacoma2015} showed that the two modes overlap and are better modeled as a probabilistic mixture. They argued that earlier evidence for discrepancy-dependent influence, such as the bounded-confidence kernel of Section \ref{model:update}, may be better explained by such a latent retention mixture.

We therefore introduce a latent response state $z_e$. At each event,
\[
z_e \sim \mathrm{Bernoulli}(\pi),
\]
where $z_e=1$ means the target is socially influenced by $s_e$, and $z_e=0$ means it retains its opinion, up to changes independent of social influence. The parameter $\pi \in [0,1]$, the \emph{influence rate}, is the probability of a social response. It is shared across agents and events.

\subsection{Response variability} \label{model:response-variability}

Given a social response, the size of the update may be very noisy. Vande Kerckhove et al.~\cite{VandeKerckhove2016} quantify the {\em intrinsic variability} of respondents using duplicated stimuli. They find that nearly two-thirds of the prediction error is due to intrinsic variability. Most studies fit models by minimizing mean squared error, which corresponds to a Gaussian likelihood with constant variance. A Gaussian likelihood is sensitive to outliers, so some studies remove them \cite{Becker2017,adams2022person,VandeKerckhove2016}. We instead model the update noise explicitly with a density $q(\cdot \mid \mu,\eta)$, where $\mu$ is the location of the update and $\eta$ controls the noise scale (the standard deviation for Gaussian noise and the mean absolute deviation for Laplace noise). Residuals are empirically heavy-tailed, so we compare Gaussian noise with heavier-tailed Laplace noise.

Many models assume that a target's opinion does not change without social influence. In real settings, exogenous factors can also move opinions and confound mechanism selection. For example, a strong exogenous drift in one direction plus weak influence in the other can look like repulsive influence if the model ignores the drift \cite{Mas2013b}. We therefore allow {\em exogenous drift}: when the target retains its opinion, its update has density $q(\cdot\mid\mu_0,\eta_0)$, where the drift $\mu_0$ is the typical opinion change absent social influence and $\eta_0$ is the scale of these changes. Setting $\mu_0=0$ recovers the common assumption that opinions change only by noise without influence.

\subsection{Opinion updates}
\label{model:update}

Since the work of Sherif \cite{sherif1936}, models have often described opinion updates after social exposure with influence-response functions, also called interaction functions or {\em interaction kernels}.
In these models, the update is the signed opinion discrepancy $\pull$ scaled by the interaction kernel $\kernel_{\param}(\glyphRadius)$:
\begin{equation}
\upd=\lambda\,\kernel_{\param}(\glyphRadius)\,\pull \,.
\label{eq:si-event-update}
\end{equation}
Here the kernel $\kernel_{\param}$, with parameters $\param$, sets the ``shape'' of the response, and the fitted coupling $\lambda\in\mathbb{R}$ sets its scale and sign. Positive values indicate attraction near zero discrepancy; negative values indicate repulsion there. 

We follow this tradition when a target is socially influenced. For event $e$, the change in opinion is
    \begin{equation}
        \updrv \mid \latent{\movind}, \pull \sim
        \begin{cases}
        \mover{q\!\left(\lambda\,\kernel_{\param}(\glyphRadius)\,\pull,\ \eta_1\right)},
          & \latent{\movind} = 1,\\[3pt]
        \stayer{q\!\left(\mu_0,\ \eta_0\right)},
          & \latent{\movind} = 0,
        \end{cases}
        \label{eq:si_emission}
    \end{equation}

where $q(\,\cdot\, \mid \mu,\eta)$ is a density with location $\mu$ and scale $\eta$. We write $\delta_e$ for a realization of the random variable $\Delta_e$ in an observed trace. A responding target's update is centered on the kernel prediction $\lambda\,\kernel_{\param}(\glyphRadius)\,\pull$ of \Cref{eq:si-event-update}, with scale $\eta_1$. A non-responding target's update is centered on a baseline drift $\mu_0$, with scale $\eta_0$.

Finally, the opinions after the event are
    \begin{equation}
        \opost = \opre + \upd \quad \text{for individual }i_e;
        \qquad
        \opost[\ell] = \opre[\ell]\quad\text{for }\ell\neq i_e.
        \label{eq:si_state_update}
    \end{equation}
In the synchronous case, we advance the configuration once all $N$ events of a round are applied. We select among five interaction kernels, four parametric and one nonparametric, described below.

\begin{figure}[t]
\centering
\includegraphics[width=\linewidth]{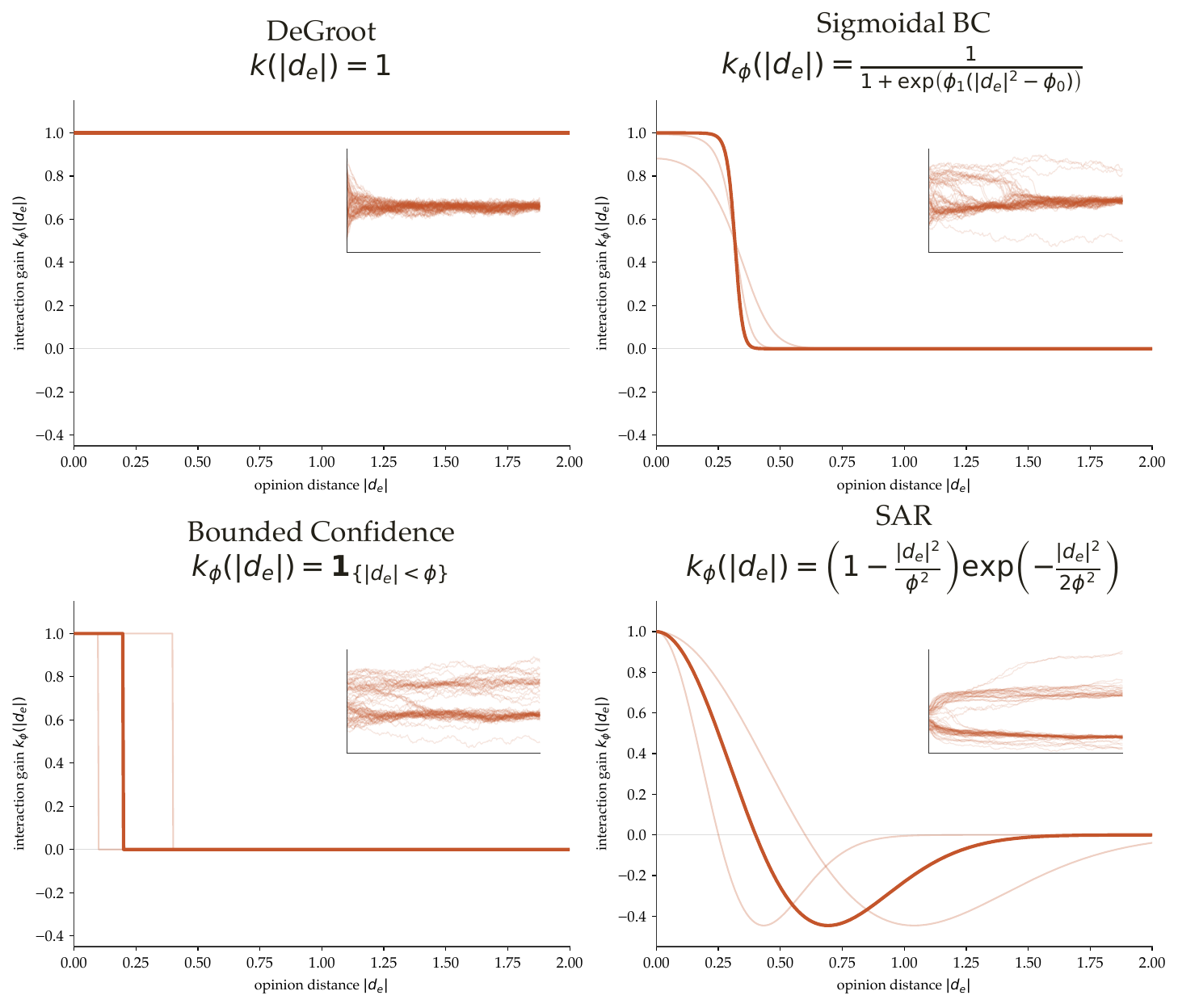}
\caption{The four parametric interaction kernels. Each panel plots the kernel $\kernel_{\param}$ against the opinion discrepancy $\lvert\pull\rvert$ between source and target at three parameter values. The value used for the inset is in bold (DeGroot has no shape parameter, so it has a single curve). Above zero the target moves toward the source, and below zero away from it; only SAR is repulsive. Insets: every node's opinion over 500 steps of the corresponding model on the Les Mis\'erables co-appearance network, with the bold parameter value, $\lambda=0.1$ and Gaussian noise of standard deviation $0.01$. DeGroot reaches consensus, bounded confidence leaves separated clusters, its sigmoidal relaxation leaves one main cluster with stragglers, and SAR splits into two groups.}
\label{fig:kernel_atlas}
\end{figure}

Different forms of $k_\phi(\glyphRadius)$ encode different hypotheses about social influence. We organize them along three contrasts: linear vs.~nonlinear, smooth vs.~non-smooth, and parametric vs.~nonparametric. Three parametric kernels are common in the opinion dynamics literature: the linear, smooth DeGroot kernel; the nonlinear, non-smooth bounded-confidence kernel; and its nonlinear smooth relaxation, the sigmoidal bounded-confidence kernel. To capture the possibility of repulsion, we add a fourth: a nonlinear smoothed attraction-repulsion kernel. \Cref{fig:kernel_atlas} plots the four parametric kernels with characteristic synthetic trajectories. Appendix~\ref{SI:model_parametric_ker} gives more detail on them.

The first formalisms, by \citet{french1956} and \citet{degroot1974}, make influence linear in the discrepancy. These {\bf French--DeGroot models} are analytically tractable, converge to consensus under broad conditions \cite{proskurnikov2017tutorial}, and are supported by a large body of empirical work \cite{VandeKerckhove2016,Takacs2016,adams2022person}.

Kernels motivated by social balance \cite{heider1946} and cognitive dissonance \cite{festinger1954} attract at small discrepancies and repel at large ones \citep{Jager_Amblard_2005,FLACHE2011,kann2023repulsiveboundedconfidencemodelopinion}. We use one of them, the {\bf smoothed attraction-repulsion (SAR) model}. Several empirical studies claim evidence of repulsion \cite{Hovland1957,Bail2018,Mazen1972,Krizan2006,Liu2015,Kozitsin2021,Tang2025b}, although methodological concerns have been raised about some of them \cite{Kozitsin2021,Mas2013b}.

{\bf Bounded-confidence (BC) models} produce dissensus without repulsion. A target is attracted to sources within a threshold and ignores the rest \citep{hegselmann2002,deffuant2000}. The \textbf{sigmoidal bounded-confidence (SBC) model}~\citep{Brooks2024} relaxes the step to a sigmoid of variable width and steepness. Empirical claims of bounded confidence \cite{moussaid2013social} may reflect confounding effects of a participant's confidence in their opinion, rather than the opinion discrepancy \cite{Chacoma2015}.

Besides these four parametric kernels, we fit a nonparametric kernel, which learns the shape of the kernel from data without assuming a functional form.
We use the Random Fourier Feature model (RFF), a memory-efficient approximation of kernel ridge regression with nonlinear feature maps.
The RFF model has the form
\begin{align*}
    \kernel_{\param}(\glyphRadius) &=
    1+\sum_{i = 1}^R \beta_i \left[\varphi_i(\glyphRadius)-\varphi_i(0)\right],\\
    \varphi_i(\glyphRadius) &= \sqrt{\frac{2}{R}}\cos(W_i\glyphRadius+b_i),
\end{align*}
where the frequencies $W_i \sim \mathcal{N}(0,\gamma^2)$ and phases $b_i \sim \mathrm{Uniform}(0,2\pi)$ are drawn once and held fixed, and the weights $\{\beta_i\}_{i=1}^R$ are learned under a ridge penalty.
Centering the features at $\glyphRadius=0$ pins $\kernel_{\param}(0)=1$. So $\lambda$ is the coupling at zero discrepancy, as for the parametric kernels, and $\beta=0$ recovers the linear DeGroot kernel.
We fix $R=256$ features. The bandwidth $\gamma$ sets how quickly the kernel can vary with $\glyphRadius$; we choose it by cross-validation with the one-standard-error rule (\Cref{sec:inference}).
\Cref{SI:model_nonparam_ker} gives more detail on fitting this kernel.

All five kernels rest on two assumptions: (1) all individuals in the population share the same kernel, and (2) the kernel depends only on the opinion discrepancy between source and target. Its absolute value sets the kernel, and its sign sets the direction of the update.
Models such as that of \citet{friedkin1990} relax the first assumption by adding individuals (``zealots'' whose opinion never changes) who behave differently from the rest of the population.
Models such as that of \citet{altafini2013} relax the second assumption by adding persistent antagonistic relationships through structural balance, rather than letting the sign of influence depend only on the opinion discrepancy.

\section{Inference methods}
\label{sec:inference}
\subsection{Likelihoods}

The generative model defines a likelihood for the observed data $\mathcal{D}$ given parameters $\Theta$. Maximum likelihood estimation seeks the parameters $\Theta$ that maximize $\probsym(\mathcal{D} \mid \mathbf{X}^{(0)}; \Theta)$, where $\mathbf{X}^{(0)}$ is the initial opinion state. We treat $\mathbf{X}^{(0)}$ as given, because the model describes only updates, not initial values.
More generally, let $\mathbf{X}^{(e)}$ be the opinion state of every agent before event $e$.

Conditioned on the opinion state before each event, the likelihood factors over events into the identity of the updating individual $i_e$ and the realized opinion update $\delta_e = x_e^+ - x_e^-$: 
\begin{align}
    \probsym(\mathcal{D} \mid \mathbf{X}^{(0)}; \Theta) = \prod_{e \in \mathcal{E}} \probsym_{\Delta, I|X}(\delta_e, i_e \mid \mathbf{X}^{(e)} ;\Theta)\;. \label{eq:complete-likelihood}
\end{align}

Taking logs gives one term per event, $\mathcal{L}_e = \log \probsym_{\Delta, I|X}(\delta_e, i_e \mid \mathbf{X}^{(e)} ;\Theta)$. 
The model has two latent variables: the response state $z_e \in \{0,1\}$ and, in networked settings, the latent source $s_e \in \srcset$. 
The joint likelihood of an update $\delta_e$ by the focal node $i_e$, given the current opinion state, is
\begin{align}
    \mathcal{L}_e
    &=\log \sum_{z_e \in \{0,1\}}\sum_{s_e \in \srcset}
    \probsym_{\Delta, I ,Z,S|X}\bigl(\delta_e, i_e, z_e, s_e \mid \mathbf{X}^{(e)}; \Theta\bigr),
\label{eqn:marginal_likelihood}
\end{align}
where $\srcset$ is the set of possible sources for event $e$. It contains either a single source or every source adjacent to $i_e$ on the graph.
\Cref{sec:si_inference} gives more detail on the likelihood.

Optimizing this marginal log likelihood \Cref{eqn:marginal_likelihood} directly is hard, because it sums over the latent assignments. We therefore use expectation maximization (EM) \citep{Dempster1977}, which alternates two steps. The E-step uses the current estimates $\Theta^{(\emiter)}$ to compute the posterior probabilities of the latent assignments. These define the expected complete-data log likelihood. The M-step maximizes this expected log likelihood over the model parameters. We repeat the two steps until convergence.
Appendix~\ref{subsec:si_em} gives the full update equations and implementation details.

\subsection{Model selection}

EM fits one candidate model. To compare candidates, we use 5-fold cross-validation. We score each fit by its held-out marginal log likelihood (MLL) per transition: for each fold, we sum over the latent variables, divide the log likelihood of the held-out opinion changes by the number of held-out transitions, and then average over folds. 
Higher scores indicate better candidate model fits.
One individual's transitions are not independent, so we block the folds by individual: all of an individual's transitions go to the same held-out set. 
Appendix~\ref{subsec:si_model_selection} gives more detail on cross-validation.

In Section~\ref{sec:results} we compare a ``ladder'' of nested model classes, each adding one mechanism to the one before. Within each class we select the model with the highest held-out MLL. We first set the \rff{} bandwidth by the one-standard-error rule \citep{hastie2009elements}: the smoothest bandwidth whose held-out MLL is within one standard error of the best. We call the inferred kernel nonlinear when the best \rff{} fit in that class has a higher held-out MLL than the best linear (DeGroot) fit. Random restarts and random-feature seeds are chosen within each fold by training MLL. To normalize for dataset size, we report every difference in nats per transition.

\subsection{Size and curvature of a kernel}
\label{subsec:size_curvature}

Beyond the linear-or-nonlinear model selection, we describe each fitted kernel by its strength and its curvature.
Both are integrals over the support of the data, $0 \le \glyphRadius \le \glyphRadius_{\max}$, where $\glyphRadius_{\max}$ is the 99th percentile of the observed discrepancies. The \emph{strength} is the root-mean-square gain
\[
S =  \left(\frac{1}{\glyphRadius_{\max}}\int_0^{\glyphRadius_{\max}} \big(\hat{\lambda}\,\gainat{\glyphRadius}\big)^2\,d\glyphRadius\right)^{1/2}.
\]
It measures the overall strength of social influence on the support of the data. The \emph{curvature} is the total variation of the gain,
\[
C = \int_0^{\glyphRadius_{\max}} \left|\hat{\lambda}\,\kernel_{\param}'(\glyphRadius)\right|\,d\glyphRadius,
\]
where the derivative $\kernel_{\param}'(\glyphRadius)$ of the \rff{} kernel is computed in closed form, mode by mode. A linear (DeGroot) kernel has a constant gain and $C = 0$. \Cref{tab:size_curvature} reports $S$ and $C$ for every dataset at every step of the ladder.

\begin{figure*}[!t]
  \centering
  \includegraphics[width=\linewidth]{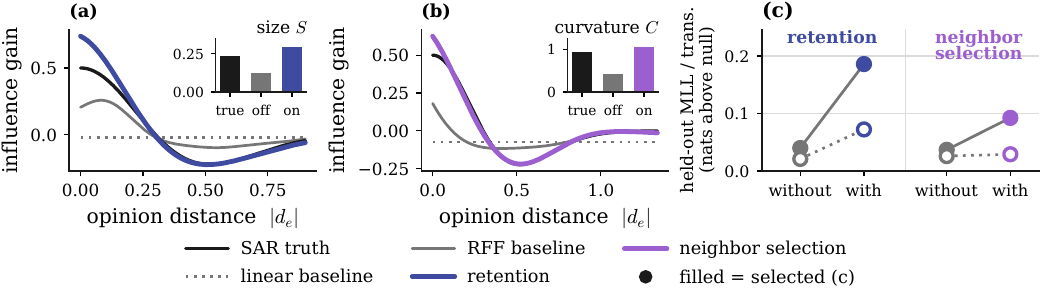}
  \caption{
    \textbf{Omitting a latent mechanism reduces the strength of the inferred kernel and
    hides the evidence for its nonlinearity.} Synthetic SAR truth with latent retention
    $\pi=0.5$, fitted without (gray) and with (color) a mechanism: retention (blue) and source selection (purple). Insets: the size $S$ (a) and the curvature $C$ (b) of the true kernel and the two fits. Omitting either mechanism roughly halves the kernel and leaves its shape close to the truth. \textbf{(c)} Held-out MLL per transition above a random-walk null, for a linear (DeGroot, dashed) and a nonparametric (RFF, solid) kernel, without (gray) and with (color) each mechanism. A filled marker is the kernel with the higher MLL.
    }
    \label{fig:synthetic_misspecification}
  \end{figure*}

\section{Results}
\label{sec:results}

\subsection{Model misspecification induces bias in synthetic data}
\label{subsec:synthetic_results}

When retention and source selection are truly present in the data-generating process, models that omit them tend to learn biased kernels. Adding a latent attention mechanism often makes nonlinearity more identifiable. We measure the effect of each mechanism with three measures: the nonlinearity verdict (whether the nonparametric kernel beats the linear kernel on held-out MLL), and the strength $S$ and curvature $C$ of the inferred kernel (\Cref{subsec:size_curvature}). 

In \Cref{fig:synthetic_misspecification}, we show results for a \sar{} kernel with latent retention at rate $\pi = 0.5$, fitted with and without each mechanism. In panel (a) we test latent retention using a single agent exposed to a single source. We first fit a linear and a nonparametric kernel without latent retention. The linear fit has $\hat\lambda < 0$, which corresponds to opinion repulsion. 
The overall strength of the nonparametric kernel without retention is roughly half of the overall strength of the true kernel but roughly the same shape. 
It is nearly indistinguishable from the linear fit: the held-out MLL difference is 0.019 nats per transition (panel (c)). With latent retention, the nonlinear fit approximately recovers the size of the true kernel, and its advantage over the linear fit grows to 0.114 nats. Under Gaussian noise, omitting latent retention reduces the kernel strength by roughly a factor of the influence rate $\pi$ (\Cref{app:omitting_mechansims}). Under Laplace noise the shrinkage is not a simple factor of $\pi$.

In panel (b) we keep retention and omit source selection on networked data, replacing the chosen source's discrepancy with the neighborhood average. The strength falls by more than half (59\%) and the curvature by about as much (55\%), resulting in a preservation of relative curvature. Thus, the averaged kernel mirrors the true kernel fit over the narrow range of discrepancies the neighborhood average reaches. This is a Berkson error \cite{Delaigle2006}: the averaged model sees the true kernel averaged over the neighborhood, and that average is pulled toward zero (Appendix~\ref{app:omitting_mechansims}). This assumes that a neighbor's deviation from the neighborhood average is unrelated to the average itself, which may not always be the case on empirical networks. Source selection helps the linear model little and the RFF a great deal. We conclude that adding latent source selection makes nonlinear and linear kernels much easier to tell apart.

In both cases, omitting a mechanism reduces the strength of the inferred kernel but preserves the shape of the true kernel. 
In general, it is more difficult to disambiguate between linear and nonlinear kernels when the kernels are smaller in overall strength because differences in likelihood will be smaller.
Under other conditions, an omitted mechanism can have the opposite effect. Omitted retention can produce spurious bounded-confidence behavior \cite{Chacoma2015}, and omitted selection can produce spurious repulsion \cite{Kozitsin2020,Mas2013b}. Either way, reliable inference of interaction kernels requires explicit modeling of the latent mechanisms by which agents respond to their information environment.

\begin{table*}[t]
\centering
\setlength{\tabcolsep}{4pt}
\footnotesize
\begin{tabular}{ll>{\raggedright\arraybackslash}p{2.4cm}>{\raggedright\arraybackslash}p{2.5cm}lr}
\toprule
\textbf{Dataset} & \textbf{A node is a...} & \textbf{Opinion variable} & \textbf{Network structure} & \textbf{Source} & \textbf{Events} \\
\midrule
\multicolumn{6}{l}{\textbf{\textit{Controlled (known source exposure)}}} \\[2pt]
\adams & Participant & Position estimate & N/A & \cite{adams2022person} & 3,262 \\
\takacs & Participant & Issue position (0--100) & N/A & \cite{Takacs2016} & 1,524 \\
\becker & Participant & Numerical estimate & Star-graph spokes & \cite{Becker2017} & 2,996 \\
\midrule
\multicolumn{6}{l}{\textbf{\textit{Controlled (latent source exposure)}}} \\[2pt]
\vkg & Player & Numerical estimate (0--100) & Complete 6-person groups & \cite{VandeKerckhove2016} & 9,499 \\
\vkc & Player & Numerical estimate (0--500) & Complete 6-person groups & \cite{VandeKerckhove2016} & 12,627 \\
\midrule
\multicolumn{6}{l}{\textbf{\textit{Political}}} \\[2pt]
\senate & US senator & Nokken-Poole score & Cosponsorship & Voteview \citep{lewis2026voteview} & 982 \\
\house & US representative & Nokken-Poole score & Cosponsorship & Voteview \citep{lewis2026voteview} & 4,010 \\
\midrule
\multicolumn{6}{l}{\textbf{\textit{Social media}}} \\[2pt]
\cmv & Commenter & Stance toward OP & Co-commenting in sub-thread & \cite{SpangherStanceModel} & 2,348 \\
\spinos & Post author & Annotated 5-point stance & Reply to a post & \cite{Sakketou2022} & 545 \\
\kozitsin & VK user & Political-stance score & Friend graph & \cite{Kozitsin2021} & 5,067 \\
\midrule
\multicolumn{6}{l}{\textbf{\textit{Prediction markets}}} \\[2pt]
\markets & Trader & Implied probability of a trade & Reply to a comment & Manifold API & 14,277 \\
\bottomrule
\end{tabular}
\caption{The 11 distinct empirical source datasets (15 panels) in the current roster, grouped by data-collection setting. Events count the update events on which each fit is scored, after exact-zero updates are removed (\cref{tab:si_panel_summary} gives the count per panel and the share removed). \takacs{} includes two similar studies. Study 1 is \takacsOne{}. Study 2 has a control arm (\takacsCtl{}) and a within-subject disliking treatment (\takacsDis{}). \adams{} tests source reliability at three levels \adamsLo{}, \adamsMid{} and \adamsHi{}.} 
\label{tab:datasets}
\end{table*}

\subsection{Modeling latent mechanisms improves out-of-sample prediction across empirical datasets}
\label{subsec:results_empirical_latent}

We expect to see misspecification effects in synthetic data generated using attention mechanisms that are not explicitly modeled. Interestingly, we see that models with latent mechanisms also predict held-out updates better in a wide range of empirical datasets (described in \Cref{tab:datasets}). Since the latent variables are not observed, a better fit is not necessarily evidence of the psychological mechanism. Our model only measures how well neighbors' opinions predict an update but cannot make causal claims about influence. Three were gathered in controlled settings where the focal agent sees a single known source (\adams{} \cite{adams2022person}, \takacs{} \cite{Takacs2016} and \becker{} \cite{Becker2017}). Two were gathered in controlled settings with a small number of sources (\vkg{} and \vkc{} \cite{VandeKerckhove2016}). Six more come from political and social media settings with multiple sources (\senate{} \cite{lewis2026voteview}, \house{} \cite{lewis2026voteview}, \markets{}, \spinos{} \cite{Sakketou2022}, \kozitsin{} \cite{Kozitsin2021} and \cmv{} \cite{SpangherStanceModel}). Several datasets contain sub-datasets. For example, \adams{} has three subsets, \adamsLo{}, \adamsMid{} and \adamsHi{}, in which the source's signaled reliability is low, medium or high (20, 50 and 80\%). \takacs{} contains two studies, which we split into three datasets. The second study uses a within-subject disliking treatment, \takacsDis{}, to investigate repulsive interactions.

To measure what each model component adds, we build a ``ladder'' of nested model classes, adding one mechanism at each step.
\begin{itemize}
  \item \textbf{Kernel only}: linear and nonlinear kernels with Gaussian noise (as least squares implicitly assumes) and no latent mechanisms.
  \item \textbf{Heavy-tailed noise}: as above, with heavy-tailed Laplace noise added as an alternative to Gaussian noise.
  \item \textbf{Latent retention}: adds latent retention, so a focal agent may not update its opinion in response to a source.
  \item \textbf{Latent source selection}: adds latent source selection, so a focal agent may not attend to all available sources in its environment.
\end{itemize}

At each step we select the model with the highest held-out MLL across all models up to that step. We select between the four parametric kernels and the nonparametric RFF described in \Cref{model:update}. In practice, when a parametric kernel is selected the shape is almost always near-linear, differing from a DeGroot kernel by a negligible amount ($<10^{-3}$ nats on average). We therefore compare only DeGroot to the flexible nonparametric RFF.

For example, the latent retention step includes all models with nonlinear kernels, heavy-tailed noise and latent retention, but no latent source selection. A step is free to leave out a mechanism that does not improve held-out MLL.
Some figures and tables number the classes as six rungs. Rung 0 is the random-walk null, and the kernel-only step is split in two: rung 1 allows only the linear kernel, and rung 2 adds the nonlinear and parametric kernels. Rungs 3, 4 and 5 are the heavy-tail, retention and source-selection steps.

\begin{figure}[p]
\centering
\includegraphics[width=\linewidth,height=0.8\textheight,keepaspectratio]{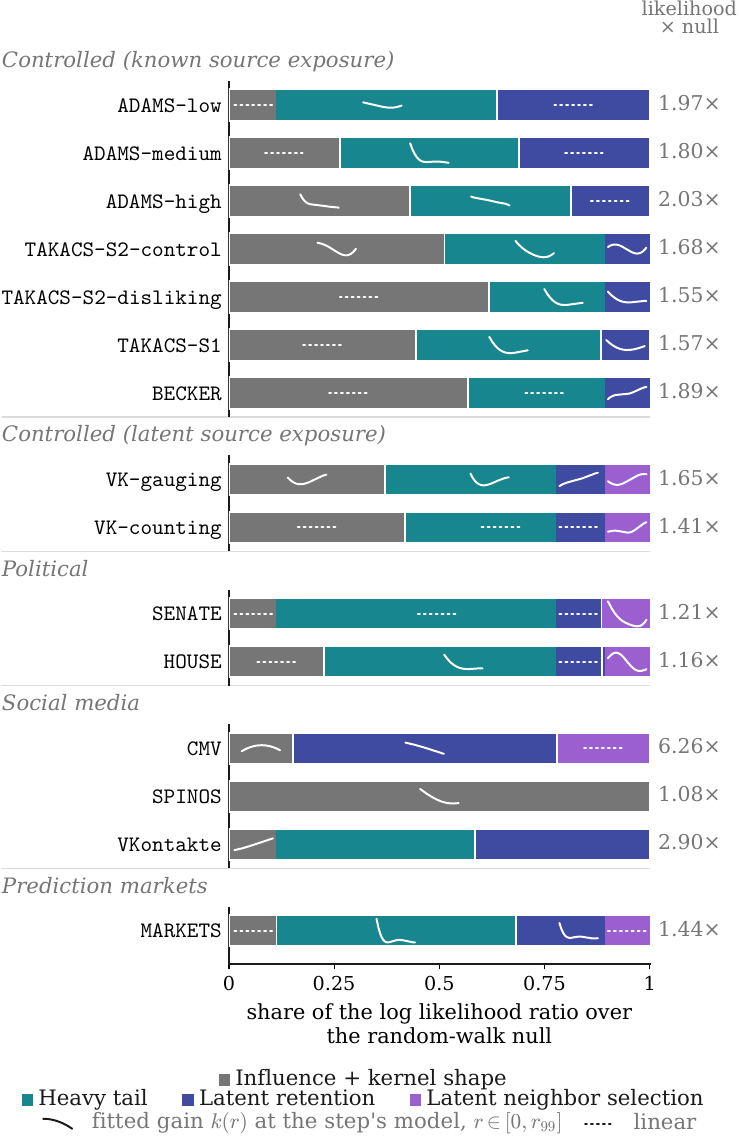}
\caption{
  \textbf{Predictive gain of each mechanism.} Each bar splits the held-out log likelihood ratio over the random-walk null into the share gained as mechanisms are added from left to right, so every bar sums to one. Each segment is the gain given the segments to its left; the split depends on this order, while the total does not. The number at its end is the total gain, as a per-transition likelihood multiplier over the null. Gray segments are the baseline; colored segments are the mechanisms proposed here. A pictogram shows the shape of the inferred kernel for each fit. Vertical scales are normalized per dataset, and kernels are centered within each bar. For \kozitsin{} the random walk null is selected on all rungs above the baseline so no pictogram is shown.
  }
\label{fig:explained_deviance}
\end{figure}

We summarize our findings in \Cref{fig:explained_deviance}. It shows the predictive gain of each mechanism as a multiple of the likelihood that a random-walk null model assigns to the data.
Bars are proportional to the marginal gain at each step of the ladder. Each step shows the predictive gain \emph{given the steps before}. A different order of steps would split the same total differently. 
Appendix~\ref{app:empirical_results} gives full results for all datasets.

\paragraph*{Opinion update noise is heavy-tailed.}

In all but one of our 15 empirical datasets, the heavy-tailed Laplace noise model is selected over the Gaussian, and it typically accounts for a large share of the predictive gain.

Heavy-tailed noise means that large opinion updates are more common than a Gaussian model predicts. Other mechanisms, such as individual differences in susceptibility \cite{friedkin1990}, may also produce heavy-tailed residuals. Our model does not distinguish between these possibilities, but it does show that the heavy-tailed Laplace distribution is a better predictive model than the Gaussian. 

The Laplace distribution also has a sharp peak at zero, consistent with many opinion updates being small or negligible.
However, this peak often cannot account for all the small updates: even after we remove exact-zero transitions, latent retention further improves predictive gain for calibrated residuals.

The noise model also changes what the kernel estimates. A Gaussian likelihood fits the conditional mean update; a Laplace likelihood fits the median. When retention is common and not modeled, the Laplace fit can collapse to near zero, and the kernel absorbs the shape of the retained updates. This increases the curvature of the fit. \Cref{tab:size_curvature} shows this: across the 13 datasets where the noise model changes, the size of the \rff{} fit falls on 11, and its curvature still rises on 8. The shrinkage is largest where retention is common, as in \adamsLo{} and \markets{}. On six datasets the heavy tail turns a linear verdict into a nonlinear one, often by a hair. On \adamsLo{}, \adamsMid{} and \markets{} the latent mechanisms turn it back, and on \takacsDis{} and \takacsOne{} it survives. \house{} goes both ways: retention turns it back, and selection makes it nonlinear again (\Cref{fig:nonlinearity_gap_scatter}).

Overall, a heavy-tailed noise model is preferred on almost every dataset. It reduces the strength of the inferred kernel and increases its curvature.

\paragraph*{Latent retention improves prediction on most datasets.}

Although the Laplace distribution can model many small updates, adding latent retention at its rung of the ladder improves held-out prediction on 13 of our 15 empirical datasets (\Cref{tab:si_ladder_compact}). It adds nothing on \senate{} and \spinos{}, and on \house{} it adds under 3\% of the predictive gain and leaves the fitted kernel unchanged. Once selection is added, retention no longer helps on \house{}, so its selected model has none (\Cref{tab:si_winners_compact}). Latent retention appears especially important when the stimuli are low-confidence.
Two caveats bear on how to interpret the retention step. First, it absorbs any update near zero, so it can stand in for heterogeneous susceptibility \cite{friedkin1990}, rounding, or disengagement as well as for retention. Second, we remove exact-zero updates before fitting, so $\hat\pi$ is the share of influence among nonzero updates. We can construe our model as a three-mode mixture with a third mode that represents exact-zero transitions. The fraction of exact-zero transitions is then the mixture weight $\pi_2$ for that mode. This fraction depends strongly on how opinions are measured: On coarse scales, such as the 100-point scale of \takacs{}, exact-zero rates are high, between 56\% and 60\%. In \adams{}, where opinions are the pixel coordinates of mouse clicks, there are no exact zeros. 
See \Cref{tab:si_panel_summary} in Appendix~\ref{app:dataset_prep} for the exact-zero fraction in each dataset. We retain the common practice of filtering exact-zero updates. 

\begin{figure*}[t]
    \centering
        \includegraphics[width=\textwidth]{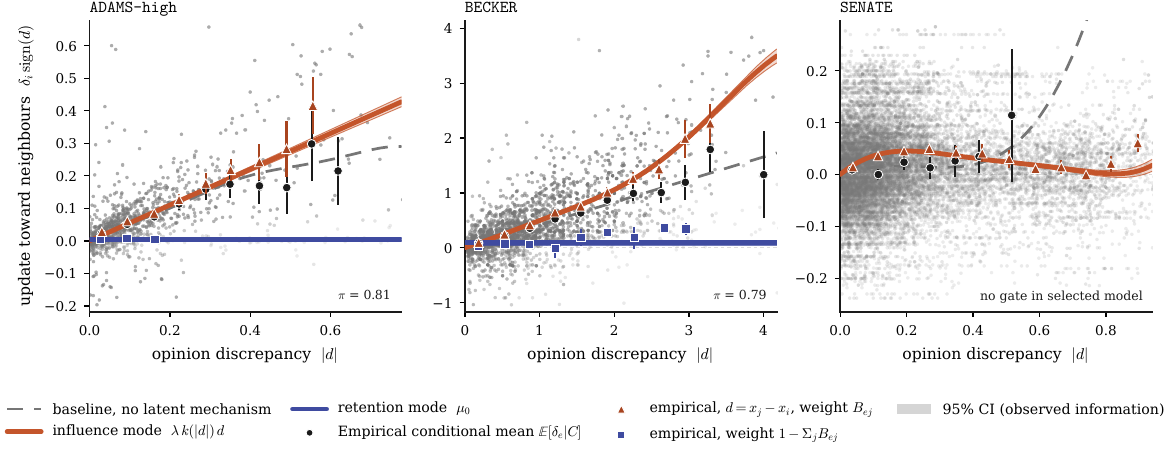}
    \caption{\textbf{The two modes of the selected model.} Update toward the source against opinion discrepancy $|\glyphPull|$. Gray points are individual updates (jittered), and black dots are binned means with 95\% intervals. Lines show the selected model's influence mode (orange), its retention mode $\mu_0$ (blue), and the fit without latent mechanisms (gray dashed). $\jointresp$ is the posterior probability that update $\ev$ was influence from source $\src$. Triangles place each update at every candidate's discrepancy $\pull[\es] = \glyphOpinion_{\glyphSource} - \opre$, weighted by $\jointresp$. Blue squares weight each update by its probability of retention, $1 - \sum_{\src}\jointresp$. On the two \vk{} games retention rescales the kernel without bending it.
    }
    \label{fig:kernel_comparison}
\end{figure*}

Retention increases the inferred size of social influence on most datasets where it is selected.  As can be seen in \Cref{fig:kernel_comparison}, the conditional median (gray dotted) falls between the medians of the influence and retention mode, mixing together updates due to influence and retention. Adding retention separates the two modes, increasing the size of social influence. Kernel size rises on 9 of these 13 datasets, as the synthetic data predict (\Cref{subsec:synthetic_results}). The exceptions are the three \takacs{} panels, where it falls slightly, and \house{}, where the fit does not change. 

Retention by itself does not change the shape of the kernel. If the chance of retaining one's opinion is independent of the opinion discrepancy, omitting retention reduces the strength of the kernel but leaves its shape unchanged. The shape changes only when retained events are either concentrated at certain discrepancies or retention is high enough to collapse the kernel to near zero. In the former case the retained updates drag the median update down at some distances in a distance-dependent way and bend the fitted kernel. In the latter case the retention rate is so high it collapses the baseline kernel to essentially zero to large distances, adding a kink in the kernel near the origin. As a result, once latent retention is added, the nonparametric kernel no longer beats the linear one (DeGroot) on any of the three \adams{} panels. 

The three \adams{} panels show both cases. How often participants respond to a source depends on how reliable it is. The \adams{} experiment provides participants with a confidence rating for each stimulus. In \adamsHi{}, where the source is reliable 80\% of the time, we estimate a high influence rate of $\hat{\pi} = 0.81$. Here, the observed fraction of retained updates increases with discrepancy: participants are more likely to retain their opinion when a stimulus is highly discrepant with their own, even if the stimulus is generally reliable. 
In \adamsLo{}, the stimulus is reliable only 20\% of the time, and the influence rate is much lower: $\hat{\pi} = 0.45$. This drags the median to nearly zero without retention. 

Adding retention removes both sources of distortion, and the bend that the nonparametric kernel had found disappears on all three \adams{} datasets. With retention DeGroot kernels are selected for all three \adams{} datasets. 

Few datasets show evidence of drift. The mean of the retention mode, $\mu_0$, is small relative to its scale in every dataset. This includes the social media datasets, where exogenous opinion shifts are plausible but a constant drift does not capture them.

The three social media datasets \cmv{}, \kozitsin{} and \spinos{} have different retention results. In \spinos{}, latent retention is not selected. The best model (a nonparametric kernel with Gaussian noise and no latent mechanism) is only marginally better than the random-walk null, so these data may carry little signal of social influence through the network we observe. In \cmv{} and \kozitsin{}, latent retention gives large predictive gains, and these are the two datasets where our richer model class improves most over the random-walk null. On \kozitsin{} that gain comes almost entirely from the heavy tail and retention. No kernel beats the random-walk null there by more than $10^{-3}$ nats per update, so we report the null as tied (\Cref{tab:si_winners_compact}).

\paragraph*{Latent source selection improves prediction over neighborhood averaging on most networked datasets.}

We select models with latent source selection (as opposed to neighborhood averaging) on 6 of 8 networked datasets, with approximate ties on the other two. Predictive gain ranges from a negligible 0.5\% of the total $\Delta$MLL over the null baseline for \markets{} to nearly 22\% in the case of \cmv{}. We read this as evidence against the neighborhood average, not as evidence that agents attend to exactly one neighbor: weighted attention, or attention to a few neighbors, might fit as well or better and is a direction for future work. 

Averaging compresses the range of observed discrepancies. A user's discrepancy with the neighborhood average can be much smaller than their discrepancy with individual neighbors. The averaged kernel is therefore fitted over a narrower range, and averaging also smooths its shape, as described in \Cref{subsec:synthetic_results}. Changes beyond that range are invisible to the averaged model.

For example, discrepancies in \house{} and \senate{} are bimodal: within-party edges tend to have small discrepancies, while between-party edges tend to have large ones. In the \senate{} panel of \Cref{fig:kernel_comparison}, the observations form clusters below $|d|<0.4$ and above $|d|>0.6$. The influence-mode fit extends beyond $|d|=0.8$, while discrepancies from the neighborhood average reach only about $|d|=0.5$. Over that compressed range, the fit without latent mechanisms suggests that influence increases with distance. A similar effect appears in \house{}. Including source selection reveals bounded-confidence-like behavior in both datasets that is hidden by neighborhood averaging.

\cmv{} is the other exception. Opinions are almost binary. The neighborhood average falls between two well separated modes. A user who adopts one neighbor's view, moving between modes, overshoots the average. The averaged kernel can only describe that with a nonlinear kernel falling with distance. With source selection, the selected kernel is linear. Source selection accounts for 22\% of the predictive gain over the random-walk null on \cmv{}, the largest share of any dataset. \house{}, \senate{} and \cmv{} show that selection becomes especially important when opinions are very polarized. All other datasets are not nearly as affected with $<4\%$ of discrepancies beyond the support of the neighbor average. 

\subsection{Latent mechanisms change the shape of interaction kernels}
\label{subsec:model_selection}

Each mechanism changes the inferred kernel in two ways: overall strength $S$, which measures the typical magnitude of the fitted gain, and curvature $C$, which measures how much that gain varies with discrepancy. Across all 15 datasets, each mechanism has a predictable effect on strength, but its effect on curvature depends on where retained updates occur across discrepancies and on the range of neighbors' opinions. A nonparametric kernel absorbs whatever part of the updates the rest of the model leaves unexplained: for retention, where the retained updates fall; for selection, how far single neighbors reach past the average.

\begin{figure}[t]
    \centering
    \iftoggle{submission}{
        \includegraphics[width=\linewidth]{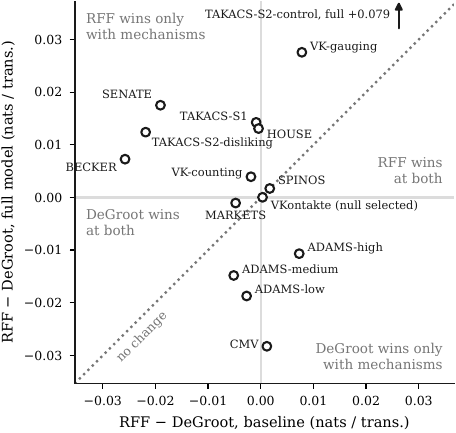}
    }{
        \includegraphics[width=0.6\linewidth]{figures/final_figures/fig_si_nonlinearity_gap_scatter.pdf}
    }
    \caption{\textbf{Where do latent mechanisms expose nonlinearity?} Difference in held-out MLL between the best nonlinear and linear kernels (RFF - DeGroot), without latent mechanisms (x-axis) and with both latent mechanisms and noise-model selection (y-axis). The upper-left quadrant holds datasets that look linear without the mechanisms but nonlinear with them; 6/15 datasets fall there. Distance from the diagonal shows the change in the MLL gap.
    }
    \label{fig:nonlinearity_gap_scatter}
\end{figure}

These changes in shape change our conclusions. \Cref{fig:nonlinearity_gap_scatter} compares the held-out advantage of the nonlinear \rff{} kernel over DeGroot without latent mechanisms and with all of them. Leaving aside \kozitsin{}, where no kernel beats the null at the top of the ladder, 5 of 14 datasets favor a nonlinear kernel without mechanisms and 9 with them. The verdict changes on eight. On six, the mechanisms reveal nonlinearity that the baseline had hidden (upper left): \becker{} through retention, \senate{}, \house{} and \vkc{} through selection, and \takacsDis{} and \takacsOne{} through the heavy tail. Two move the other way, \adamsHi{} and \cmv{}, the cases discussed above: retention (\adams{}) or selection (\cmv{}) removes curvature that the baseline had absorbed. In the \takacs{} experiments, retention flattens the kernel without restoring a linear verdict.

The latent mechanisms do not resolve the empirical disagreement about nonlinearity, but they highlight a source of this disagreement. They improve held-out prediction and can shift conclusions about mechanisms. Models of social influence should therefore account for how and when individuals pay attention, and to what.

\begin{table}[t]
  \centering
  \small
  \begin{tabular}{lcccccccc}
\toprule
 & \multicolumn{2}{c}{kernel only} & \multicolumn{2}{c}{+ heavy tail} & \multicolumn{2}{c}{+ retention} & \multicolumn{2}{c}{+ selection} \\
\cmidrule(lr){2-3}\cmidrule(lr){4-5}\cmidrule(lr){6-7}\cmidrule(lr){8-9}
panel & $S$ & $C$ & $S$ & $C$ & $S$ & $C$ & $S$ & $C$ \\
\midrule
\multicolumn{9}{l}{\textit{Controlled (known source exposure)}} \\
\adamsLo{} & 0.13 & 0.062 & 0.029 & 0.0097 & 0.20 & 0.24 &  &  \\
\adamsMid{} & 0.25 & 0.21 & 0.16 & 0.34 & 0.40 & 0.28 &  &  \\
\adamsHi{} & 0.49 & 0.52 & 0.46 & 0.24 & 0.59 & 0.40 &  &  \\
\takacsCtl{} & 0.33 & 0.45 & 0.33 & 0.61 & 0.31 & 0.29 &  &  \\
\takacsDis{} & 0.35 & 0.41 & 0.36 & 0.59 & 0.35 & 0.32 &  &  \\
\takacsOne{} & 0.43 & 0.63 & 0.41 & 0.69 & 0.38 & 0.30 &  &  \\
\becker{} & 0.43 & 0.23 & 0.40 & 0.27 & 0.58 & 0.49 &  &  \\
\addlinespace
\multicolumn{9}{l}{\textit{Controlled (latent source exposure)}} \\
\vkg{} & 0.44 & 0.40 & 0.39 & 0.52 & 0.55 & 0.55 & 0.64 & 0.58 \\
\vkc{} & 0.29 & 0.038 & 0.22 & 0.15 & 0.27 & 0.079 & 0.40 & 0.36 \\
\addlinespace
\multicolumn{9}{l}{\textit{Political}} \\
\senate{} & 0.12 & 0.31 & 0.077 & 0.21 & 0.17 & 0.25 & 0.18 & 0.59 \\
\house{} & 0.16 & 0.090 & 0.14 & 0.17 & = & = & 0.13 & 0.28 \\
\addlinespace
\multicolumn{9}{l}{\textit{Social media}} \\
\cmv{} & 0.83 & 0.33 & = & = & 2.1 & 1.5 & 1.1 & 0.20 \\
\spinos{} & 0.59 & 0.72 & = & = & = & = & = & = \\
\kozitsin{} & 0.026 & 0.020 & $<$0.001 & 0.0012 & 0.0022 & 0.0020 & 0.0026 & 0.0025 \\
\addlinespace
\multicolumn{9}{l}{\textit{Prediction markets}} \\
\markets{} & 0.23 & 0.41 & 0.072 & 0.24 & 0.61 & 1.2 & 0.73 & 0.94 \\
\bottomrule
\end{tabular}

  \caption{\textbf{Size and curvature of the \rff{} kernel up the ladder.} Size $S$ is the root-mean-square gain and curvature $C$ its total variation, both on the support of the data (\Cref{subsec:size_curvature}). Each column pair adds one step: the kernel alone, heavy-tailed noise, latent retention, and latent source selection. Each cell is the best \rff{} fit at that step, regardless of whether or not it is selected over DeGroot.  ``='' marks a step at which the best \rff{} fit does not change. Dyadic datasets have no selection step. At the selection step the kernel is evaluated over single-neighbor discrepancies, a longer range than at earlier steps; Appendix~\ref{app:selection_support} compares the two on their shared range. Values are means over CV folds. Appendix~\ref{app:mechanism_evidence} gives the data behind the retention and selection columns.}
  \label{tab:size_curvature}
\end{table}

\subsection{Exploring learned kernels}
\label{subsec:results_kernels}

We find no evidence of repulsion that depends on opinion discrepancy. Where the \sar{} kernel is selected, its repulsive regime lies beyond the observed discrepancies.
In every dataset we studied, the fitted interaction kernel is attractive across the observed range of opinion discrepancies.
Our kernels are shared by the whole population and depend only on the discrepancy, so this does not rule out repulsion that is specific to some identities or topics \cite{Bail2018,Flache2017}.
\Cref{fig:kernel_comparison} shows a subset of datasets. It plots the conditional mean of the best baseline fit (no latent mechanisms), and the conditional means of the retention and influence modes of the selected model.

Liu and Srivastava \cite{Liu2015} find evidence of repulsive dynamics using congressional voting data. In contrast to \senate{} and \house{}, they build the social network by including seating arrangements and committee co-membership as proxies for social influence. This suggests that the choice of network, i.e., which edges carry social influence, may change conclusions about the interaction kernels governing social influence.

\section{Discussion}
\label{sec:discussion}

We have proposed a probabilistic generative agent-based model (PGABM) that explicitly models several commonly hypothesized or empirically observed mechanisms: social influence, intrinsic variability, latent retention, and latent source selection.
The model reflects a common situation: a researcher can observe or control an individual's information environment, but not the individual's attention to it. Latent retention and source selection model this attention.
We fit the model efficiently with expectation maximization, and we use it to select among a suite of interaction kernels and noise models.
To our knowledge, this work is the first to compare several families of interaction kernels across many empirical datasets with latent attention mechanisms.

We have argued that omitting latent retention and source selection can bias inference of social influence.
On synthetic data, omitting either latent mechanism reduces the overall strength of the inferred kernel, which obscures the evidence for its nonlinearity.
On a wide range of empirical datasets, modeling attentional mechanisms improves held-out predictive likelihood on most datasets. When we add these latent mechanisms into our model selection process as described in \Cref{subsec:results_empirical_latent}, they add more to prediction than the choice of parametric kernel family.
In contrast to some earlier hypotheses \cite{Kozitsin2021,Mas2013b}, modeling latent mechanisms does not simply remove nonlinearity from inferred kernels. Adding heavy-tailed noise and latent mechanisms tends to increase the inferred effect of social influence. It changes the determination of whether the interaction kernel governing social influence is linear or nonlinear, and we see nonlinear kernels selected more often. Each mechanism changes the kernel's overall strength in a predictable fashion. When the heavy-tailed Laplace noise is selected, the resulting kernel has smaller overall strength and larger curvature. Retention enlarges the overall strength of the kernel (on 9 of 13 datasets). Selection changes its curvature in a direction set by how far single neighbors sit from the neighborhood average. 
Modeling intrinsic variability explicitly also matters. A heavy-tailed noise model is preferred on nearly every dataset. 

Our evidence has limits. It is predictive: the latent variables are not observed, so a better held-out fit supports a model with retention and selection without identifying them as psychological mechanisms. The retention mechanism in our model can absorb other sources of near-zero updates, and we remove exact zeros before fitting, so $\hat\pi$ describes nonzero updates only. We hold $\pi$ fixed across discrepancies, and where retention depends on discrepancy some of that dependence can be explained with an appropriate choice of interaction kernel. On the observational datasets, neighbors' opinions can predict an update through homophily, shared external shocks, or tie formation in addition to influence. Our congressional network pools all twelve Congresses, so later ties could shape the exposure of earlier transitions. Our results there describe predictability given the network rather than causal influence.

There are several exciting directions for future work that builds on our framework.
We considered four parametric families and one nonparametric family of interaction kernels, but many others are possible.
Kernels that are not translation invariant could be especially interesting. Each of our kernels depends on the opinions of the focal agent $x_{i}$ and source $x_{s}$ only through the difference $x_{s}-x_{i}$. However, the influence of a source may depend on a more general function $f(x_{i},x_{s})$ of the two opinions; neural network kernels are one natural candidate for this generalization.
Noise models beyond the Gaussian and Laplace families could also be considered.
Another important direction is richer models of source selection. Selection could depend on the opinions of the focal agent and candidate sources, on an external measure of the sources' influence or prestige, or on the previous selections of other agents. An agent could also weigh several neighbors at once, and the retention rate could depend on the discrepancy.

Our results also point to future experiments.
Across empirical datasets, our pipeline selects quite different latent retention rates and kernel shapes.
Controlled experiments could isolate and test the mechanisms we propose, especially latent retention and source selection.
We hope that such studies, combined with extensions of our framework, can help unite model-based and empirical understanding of social influence and opinion dynamics.

\FloatBarrier

\section*{Data Availability}
All code needed to reproduce the results of the paper is available in the \href{https://github.com/WillHWThompson/Learning-Opinions-With-Latent-Attention}{project repository}.

\textbf{AI Use Disclaimer} During the preparation of this manuscript, the authors used Claude Opus 5.5 to assist with coding. The authors reviewed and edited code for accuracy and assume full responsibility for the results. Generative AI was not used for generating text, and was used only in a limited capacity for grammatical edits.  

\section*{Acknowledgments}
We are grateful to Thomas Gebhart for early contributions and helpful discussions.
This group was convened as part of the 2023 Mathematics Research Community on ``Complex Social Systems;'' we are grateful to the American Mathematical Society, the National Science Foundation, and the organizers.
PSC acknowledges support from the National Science Foundation under awards DMS-2407058 and 1827373.
HZB acknowledges support from the National Science Foundation under awards DMS-2109239 and MRI-2408259.

\bibliographystyle{unsrtnat}
\bibliography{references}

\appendix

\section{Generative model}\label{sec:si_model}

Many controlled laboratory experiments isolate the effect of a single stimulus on a participant's opinion. A participant is asked to estimate their opinion, either in the form of a numerical-estimation task \cite{VandeKerckhove2016,Becker2017,moussaid2013social,Chacoma2015} or through the quantification of a subjective opinion on a numerical scale \cite{Takacs2016,Mas2013b,Friedkin2021,Hovland1957}. The participant is then exposed to new information: the opinion of another participant \cite{Takacs2016,moussaid2013social,Chacoma2015,adams2022person}, a predetermined false signal \cite{Takacs2016,asch1951}, or an article intended to persuade the participant \cite{Wood2019,Nyhan2010,Hovland1957}. After the exposure, the participant is again solicited for their opinion, and the difference before and after the stimulus is the response $\upd$.

Because both $\pull$ and $\upd$ are directly measured in source-identified exposure studies, the functional form of the gain $k_{\param}(\glyphRadius)$ is recoverable by plotting $\pull$ against $\upd$. Methods employ linear regression (sometimes with polynomial terms) \cite{adams2022person,Wood2019,Mas2013b}, a nonparametric method such as LOWESS \cite{Takacs2016}, binned curves \cite{Kozitsin2021}, decision trees \cite{moussaid2013social}, or influence maps \cite{Chacoma2015}.

\subsection{Parametric kernels}
\label{SI:model_parametric_ker}

Here we present mathematical definitions of the four parametric kernels. Each is a map $k_{\param}: \mathbb{R}_{\ge 0} \rightarrow \mathbb{R}$ of the opinion distance $\glyphRadius$, where $\param \in \Omega$ and $\Omega$ is the parameter space of the kernel family.

\subsubsection{DeGroot: linear assimilation}

In the classical DeGroot model~\cite{degroot1974}, a target updates its opinion as a weighted average of the source's opinions. Taking equal weights and dyadic interactions, this reduces to a constant kernel:
\begin{equation}
    \label{eq:degroot-kernel}
    \kernel(\glyphRadius) = 1
\end{equation}
The family has no shape parameter: the strength of social influence is carried entirely by the coupling $\lambda$, and a free constant here would be confounded with it (only the product would be identified), so we fix it at one. Within this model, opinions tend to converge to a single value (consensus) under relatively broad settings.

\subsubsection{Smoothed attractive-repulsive kernel}\label{SI:SAR}
A second modeling tradition treats distant opinions not as ignored but as repulsive: targets move away from sufficiently dissimilar sources. Bounded-confidence-with-contrast rules formalize this as social judgment's latitude of rejection~\cite{Jager_Amblard_2005}, and related models introduce negative influence or differentiation from dissimilar others~\cite{FLACHE2011}. Some DeGroot-type backfire constructions keep opinions in a compact interval by clipping, introducing nonsmoothness~\cite{Chen2021}. Inspired by the Ricker wavelet~\cite{Ricker1953}, we instead propose a smoothed attraction--repulsion kernel in which the sign of $\kernel_{\param}$ flips with opinion distance:
\begin{equation}
    \label{eq:sar-kernel}
    \kernel_{\param}(\glyphRadius)=\left(1-\frac{\glyphRadius^2}{\scalarparam^2}\right)\exp\!\left(-\frac{\lvert \pull \rvert^2}{2\scalarparam^2}\right), \qquad \scalarparam>0
\end{equation}
where $\param \in \Omega = \mathbb{R}_{>0}$ controls the scale at which attraction transitions to repulsion. In particular, the influence is assimilative for $\glyphRadius<\scalarparam$ and repulsive for $\glyphRadius>\scalarparam$.

\subsubsection{Bounded confidence kernel}



Bounded-confidence models formalize the idea that the target only assimilates opinions within a latitude of acceptance, ignoring sufficiently distant ones~\cite{hegselmann2002,deffuant2000}. The Hegselmann--Krause kernel encodes this as hard thresholding $\glyphRadius<\scalarparam$ in opinion distance:
\begin{equation}
    \label{eq:hk-kernel}
    \kernel_{\param}(\glyphRadius)=\mathbf{1}_{\{\glyphRadius<\scalarparam\}}
\end{equation}
where $\param \in \Omega = \mathbb{R}_{>0}$ is the confidence threshold: a target is influenced by a source only if their opinion distance is less than $\param$. Unlike DeGroot dynamics, bounded-confidence models can yield consensus, fragmentation, or polarization. 

\subsubsection{Sigmoidal bounded-confidence}

The hard HK threshold is an idealization of selective influence. The sigmoidal bounded-confidence model~\cite{Brooks2024} replaces the discontinuous step with a smooth cutoff, with parameters $\param = (\glyphWidth, \glyphSteep) \in \Omega = \mathbb{R} \times \mathbb{R}$. Here, positive $\glyphWidth$ sets the squared confidence radius and $\glyphSteep$ the sharpness, allowing for smooth interpolation between near-consensus and hard bounded-confidence regimes:
\begin{equation}
    \label{eq:sbc-kernel}
    \kernel_{\param}(\glyphRadius)=\frac{1}{1+\exp\!\bigl(\glyphSteep(\glyphRadius^2-\glyphWidth)\bigr)},
    \qquad \param=(\glyphWidth,\glyphSteep).
\end{equation}
Large positive~$\glyphSteep$ approximates an HK-like threshold near $\glyphRadius=\sqrt{\glyphWidth}$ when $\glyphWidth>0$; small~$\glyphSteep$ yields near-constant gain as in DeGroot. We fit both parameters without a sign constraint. A negative width with positive steepness puts the gain below one half at every distance, so the coupling at zero discrepancy is $\lambda\kernel_{\param}(0)$ rather than $\lambda$ (the SBC fit on \house{} has $\kernel_{\param}(0) \approx 0.19$). A negative steepness gives a gain that rises with distance (\becker{}).

\subsection{Learning the RFF kernel} \label{SI:model_nonparam_ker}

The RFF kernel is linear in its weights $\beta$, so its parameter step (\cref{eq:em-phi}) is a weighted ridge regression. Write $d_{s,e} = x_s - x_{i_e}$. The influence arm's location is $\lambda d_{s,e}\bigl[1 + \sum_i \beta_i (\varphi_i(|d_{s,e}|) - \varphi_i(0))\bigr]$, so the residual $\delta_e - \lambda d_{s,e}$ is linear in $\beta$ with design vector $z_{s,e} = \lambda d_{s,e}\,[\varphi(|d_{s,e}|) - \varphi(0)] \in \mathbb{R}^R$. Under Gaussian noise the step is exact:
\begin{equation*}
    \hat\beta = \Bigl(\sum_{e,s} y^{(\tau)}_{1,s,e}\, z_{s,e} z_{s,e}^\top + \rho I\Bigr)^{-1} \sum_{e,s} y^{(\tau)}_{1,s,e}\, z_{s,e}\,\bigl(\delta_e - \lambda d_{s,e}\bigr),
\end{equation*}
with an absolute ridge $\rho = 10^{-3}$. Under Laplace noise we use the same least-squares step as a surrogate. It is not the conditional maximizer of the Laplace $Q_3^{(\tau)}$, so this step can lower the likelihood; each EM run returns its best iterate. The frequencies are drawn as $W_i \sim \mathcal{N}(0, \gamma^2)$ with $R = 256$; the bandwidth grid, the four feature seeds and the one-standard-error rule are given in \cref{subsec:si_model_selection}.

\section{Inference methods}
\label{sec:si_inference}
\subsection{Likelihood}

    We aim to maximize the likelihood given by \Cref{eq:complete-likelihood}. 
    To do so, we need to explicitly describe the likelihood of a single transition in an event, including the latent response state $z_e \in \{0,1\}$ and the latent source $s_e \in \srcset$, where $\srcset$ is the set of candidate sources for event $e$. 
    We can factor this likelihood as 
    \begin{align}
        \probsym_{\Delta, I ,Z,S|X}\bigl(\delta_e, i_e, z_e, s_e \mid \mathbf{X}^{(e)}; \Theta\bigr) = \\
        \probsym_{\Delta | I, Z,S,X}\bigl(\delta_e \mid i_e, z_e, s_e , \mathbf{X}^{(e)}; \Theta\bigr) \probsym_{I}(i_e) \probsym_{S|I}\bigl(s_e|i_e\bigr) \probsym_{Z}(z_e;\Theta)\;. 
    \end{align}
   Here we have taken advantage of the fact that the choice of updating agent $i_e$ and response state $z_e$ are independent of all variables including the current state $\mathbf{X}^{(e)}$, while the identity of the source $s_e$ depends only on $i_e$. 
    Following the model description in the main text, we have 
    \begin{align}
        \probsym_Z(z_e;\Theta) &= \pi^{z_e} (1-\pi)^{1-z_e} \label{eq:likelihood-1}\\ 
        \probsym_{S|I}(s_e|i_e) &= \frac{1}{|\srcset|} \label{eq:likelihood-2}\\ 
        \probsym_I(i_e) &= \frac{1}{N} \label{eq:likelihood-3}\\ 
        \probsym_{\Delta | I, Z,S,X}\bigl(\delta_e \mid i_e, z_e, s_e , \mathbf{X}^{(e)}; \Theta\bigr) &= \\ 
        & \hspace{-6em} \begin{cases}
            q_0(\delta_e; \mu_0, \eta_0)&\quad z_e = 0 \\ 
            q_1(\delta_e; \lambda k(x_{i_e}, x_{s_e}; \phi), \eta_1) &\quad z_e = 1\;,
        \end{cases}\label{eq:likelihood-4}
    \end{align}
    where $q_0$ is the latent retention distribution with parameters $\mu_0$ (location) and $\eta_0$ (scale) and $q_1$ is the influence distribution with parameters $\lambda k(x_{i_e}, x_{s_e})$ and $\eta_1$. 
    The influence kernel $k(x_{i_e}, x_{s_e})$ may have additional parameters which we leave implicit. Throughout, the two-argument kernel includes the discrepancy: $k(x_{i_e}, x_s; \phi) = k_\phi(|d_{s,e}|)\, d_{s,e}$ with $d_{s,e} = x_s - x_{i_e}$, where the one-argument $k_\phi$ is the gain of \Cref{sec:si_model}. 
    As options for $q_0$ and $q_1$, we consider Gaussian and Laplace distributions and select between them using model selection.  
    We learn the parameters $\mu_0$, $\eta_0$, $\eta_1$, and the implicit parameters of the influence kernel via expectation maximization.

\subsection{Generalized expectation maximization}
\label{subsec:si_em}

\begin{algorithm}[t]
\caption{Generalized expectation maximization for latent influence status and pairwise attention.}
\label{alg:si_em}
\begin{algorithmic}[1]
\Require $\mathcal{D}$, candidate source sets $\srcset$, $\Theta^{(0)}$, tolerance $\epsilon$
\Ensure Fitted parameters $\hat\Theta$
\State $\emiter\gets 0$
\Repeat
    \State \textbf{E step}
    \For{each event $e$, possible state $z\in \{0,1\}$, and candidate $\src \in \srcset$}
        \State Compute $y^{(\tau)}_{z, s, e}$ via \cref{eq:em-y}
    \EndFor
    \State \textbf{M step}
    \State  $\Theta^{(\tau + 1)} \gets \left(\hat{\pi}^{(\tau+1)}, \hat{\mu}_0^{(\tau+1)}, \hat{\eta}_0^{(\tau+1)}, \hat{\param}^{(\tau+1)}, \hat{\lambda}^{(\tau+1)}, \hat{\eta}_1^{(\tau+1)}\right)$ using \cref{eq:em-q1,eq:em-q2,eq:em-q3}
    \State $\emiter\gets\emiter+1$
\Until{the relative change in $\log \probsym(\mathcal{D}\mid\Theta^{(\tau)})$ stays below $\epsilon$ (\cref{subsec:si_em_training})}
\State \Return the iterate $\hat{\Theta}^{(\tau)}$ with the highest $\log \probsym(\mathcal{D}\mid\Theta^{(\tau)})$
\end{algorithmic}
\end{algorithm}

We employ a generalized expectation-maximization algorithm (GEM) \cite{Dempster1977,wu1983convergence}. 

In the \emph{E-step} of this algorithm, we compute a probability distribution over the latent states $(z_e, s_e)$ for each event, while in the \emph{M-step} we use this distribution to compute updated estimates of our model parameters. Unlike standard EM, GEM's M-step does not need to maximize the objective function, only to increase it.

Any M-step that increases the objective cannot decrease the marginal log-likelihood $\log \probsym(\mathcal{D}|\Theta)$\cite{Dempster1977}. Some of our M-steps are gradient steps that are not guaranteed to increase the objective, so we keep the iterate with the highest marginal log-likelihood rather than the last one. We alternate the two steps until the relative improvement in the marginal log-likelihood is smaller than a predefined tolerance $\epsilon$. 

We now describe the mathematical steps of this algorithm. 
We begin with an initial estimate $\hat{\Theta}^{(0)}$ of the parameter $\Theta$, and seek to return an estimate $\hat{\Theta}^{(\tau)}$ after some number $\tau$ of iterations such that $\hat{\Theta}$ maximizes the marginal log-likelihood. 

For each event $e \in \mathcal{E}$, define the conditional distribution $r_{Z, S|\Delta, I, X}$ over the latent states $z_e$ (receptiveness to influence) and $s_e$ (information source) given the observed opinion update $\delta_e$, identity of updating agent $i_e$, and prior opinion state $\mathbf{X}^{(e)}$. 
Using Bayes' rule, we can compute 
\begin{align}
    &r_{Z, S|\Delta, I, X}(z_e, s_e|\Delta_e, i_e, \mathbf{X}^{(e)}; \Theta) = \\ 
    &\quad \quad \quad  \frac{\probsym_{\Delta, Z,S|I,X}\bigl(\delta_e,  z_e, s_e \mid i_e, \mathbf{X}^{(e)}; \Theta\bigr)}{\sum_{z', s'}\probsym_{\Delta, Z,S |I,X}\bigl(\delta_e, z', s' \mid  i_e, \mathbf{X}^{(e)}; \Theta\bigr)}\;.
\end{align}
At algorithmic timestep $\tau$, let 
\begin{align}
    y_{z, s, e}^{(\tau)} = r_{Z, S|\Delta, I, X}(z, s|\Delta_e, i_e, \mathbf{X}^{(e)}; \Theta^{(\tau)})\;. \label{eq:em-y}
\end{align}

We then use $y_{z, s, e}^{(\tau)}$ to define the objective function used in the M-step: 
\begin{align}
    Q^{(\tau)}(\Theta) = \sum_{\substack{e \in \mathcal{E} \\ z \in \{0,1\} \\ s \in \srcset}} y_{z, s, e}^{(\tau)} \log \probsym_{\Delta, Z, S | I,  X}(\delta_e,z, s|   i_e, \mathbf{X}^{(e)};\Theta) + K \label{eq:EMQ}
\end{align}
where $K = \sum_{\substack{e \in \mathcal{E} \\ z \in \{0,1\} \\ s \in \srcset}} y_{z, s, e}^{(\tau)} \log \probsym_{I | X}(i_e| \mathbf{X}^{(e)};\Theta)$ may be treated as a constant since the selection of the updating agent $i_e$ does not depend on $\Theta$. 
This function is the expected complete log-likelihood of the observed opinion increments and latent variables, with expectation over the latent states taken with respect to the distribution $y_{z, s, e}^{(\tau)}$. 
The M-step computes the next parameter estimate as 
\begin{align}
    \hat{\Theta}^{(\tau + 1)} = \argmax_{\Theta} Q^{(\tau)}(\Theta)\;. 
\end{align}
In practice, this involves several separate substeps for each of the different parameter blocks contained in $\Theta$. 
We can obtain parameter estimates by inserting \Cref{eq:likelihood-1,eq:likelihood-2,eq:likelihood-3,eq:likelihood-4} into \Cref{eq:EMQ} and maximizing with respect to the associated parameters.

\subsubsection{Influence Probability \texorpdfstring{$\hat{\pi}$}{pi}}

To compute the estimate $\hat{\pi}^{(\tau+1)}$ of the influence probability, we note that the only terms in \Cref{eq:EMQ} which depend on $\pi$ are 
\begin{align}
    Q_1^{(\tau)}(\pi) = \sum_{\substack{e \in \mathcal{E} \\ z \in \{0,1\} \\ s \in \srcset}} y_{z,s,e}^{(\tau)} \bigl( z \log \pi + (1 - z) \log (1- \pi)\bigr)\;. 
\end{align}
The estimation problem is to solve 
\begin{align}
    \hat{\pi}^{(\tau + 1)} = \argmax_{\pi} Q_1^{(\tau)}(\pi)\;. \label{eq:em-q1}
\end{align}

Differentiating with respect to $\pi$ and solving the critical point equation $\frac{\partial Q_1^{(\tau)}(\pi)}{\partial \pi} = 0$ yields 
\begin{align}
    \hat{\pi}^{(\tau+1)} = \frac{1}{\abs{\mathcal{E}}}\sum_{\substack{e \in \mathcal{E} \\ z \in \{0,1\} \\ s \in \srcset}} y_{z, s,e}^{(\tau)} z\;, 
\end{align}
i.e. $\hat{\pi}^{(\tau+1)}$ is the expected frequency of influence events with respect to the distribution $y_{z, s,e}^{(\tau)}$. 

\subsubsection{Retention arm parameters}

We now consider the location parameter $\mu_0$ and scale parameter $\eta_0$ of the retention arm. 
The corresponding term of the objective function \Cref{eq:EMQ} which depends on these parameters is 
\begin{align}
    Q_2^{(\tau)}(\mu_0, \eta_0) = \sum_{\substack{e \in \mathcal{E} \\  s \in \srcset}} y_{0,s,e}^{(\tau)} \log q_0(\delta_e; \mu_0, \eta_0)\;.
\end{align}

We aim to solve 
\begin{align}
    (\hat{\mu_0}^{(\tau + 1)}, \hat{\eta_0}^{(\tau+1)}) =  \argmax_{\mu_0, \eta_0} Q_2^{(\tau)}(\mu_0, \eta_0)\;. \label{eq:em-q2}
\end{align}

The optimizing formulae depend on the noise model $q_0$.
In the case of the Gaussian noise model 
\begin{align}
    q_0(\delta; \mu, \eta) = \frac{1}{\eta \sqrt{2\pi}} \exp \left[ - \frac{1}{2} \left(\frac{\delta - \mu}{\eta}\right)^2 \right]\;,
\end{align}
we obtain 
\begin{align}
    Q_2^{(\tau)}(\mu_0, \eta_0) = \sum_{\substack{e \in \mathcal{E} \\  s \in \srcset}} y^{(\tau)}_{0,s,e} \left[-\frac{1}{2\eta_0^2}(\delta_e - \mu_0)^2 - \log \eta_0 \right] + C\;.
\end{align}
where $C$ does not depend on the parameters. 

The optimization in $\mu_0$ is a weighted least-squares problem with solution 
\begin{align}
    \hat{\mu}_0^{(\tau + 1)} = \frac{\sum_{e, s} y^{(\tau)}_{0, s,e}\,\delta_e}{\sum_{e, s} y^{(\tau)}_{0, s,e}}\;, \label{eq:em-mu0-gauss}
\end{align}
i.e. $\hat{\mu}_0^{(\tau + 1)}$ is the expected mean update in retention events, with expectation taken with respect to  $y^{(\tau)}_{0, s,e}$. 
The optimization in $\eta_0$ gives 
\begin{align}
    \hat{\eta_0}^{(\tau + 1)} = \sqrt{\frac{\sum_{e, s} y^{(\tau)}_{0,s,e} \bigl(\delta_e - \hat{\mu}_0^{(\tau+1)}\bigr)^2}{\sum_{e, s} y^{(\tau)}_{0,s,e}}} \;. \label{eq:em-eta0-gauss}
\end{align}


In the case of the Laplace noise model 
\begin{align}
    q_0(\delta; \mu, \eta) = \frac{1}{2\eta} \exp\left(-\frac{|\delta - \mu|}{\eta}\right)\;.
\end{align}

For Gaussian noise, the location problem is a weighted least-squares problem, the minimizer of which is the weighted empirical mean, and the scale is set by the weighted empirical standard deviation. For Laplace noise, the least-squares problem is replaced with a weighted absolute-deviation problem, the minimizer of which is the weighted median, and the scale is set by the weighted empirical mean absolute deviation. 
\Cref{eq:em-mu0-gauss,eq:em-eta0-gauss} can then be rewritten for Laplace noise:
\begin{align}
    \hat{\mu}_0^{(\tau + 1)} &= \operatorname*{weighted\,median}_{e}\Bigl(\{\delta_e\}, \bigl\{\textstyle\sum_s y^{(\tau)}_{0,s,e}\bigr\}\Bigr)\;, \label{eq:em-mu0-lap}\\
    \hat{\eta}_0^{(\tau + 1)} &= \frac{\sum_{e, s} y^{(\tau)}_{0,s,e} \bigl|\delta_e - \hat{\mu}_0^{(\tau+1)}\bigr|}{\sum_{e, s} y^{(\tau)}_{0,s,e}}\;. \label{eq:em-eta0-lap}
\end{align}

\subsubsection{Influence arm parameters}

We now turn to the influence arm. 
Here we must estimate the noise $\eta_1$, as well as the parameters $\phi$ of the influence $\lambda k(x_{i_e}, x_s;\phi)$. 
Every kernel's gain except SBC's is normalized to one at zero discrepancy, $k_\phi(0) = 1$. This fixes the scale of $k$ and makes $\lambda$ and $\phi$ identifiable. The SBC gain at zero is $k_\phi(0) = 1/(1+e^{-\phi_0\phi_1})$, which is close to one only when $\phi_0\phi_1$ is large and positive, so for SBC the coupling at zero discrepancy is $\lambda k_\phi(0)$ rather than $\lambda$.

The term that depends on these parameters is 
\begin{align}
    Q_3^{(\tau)} (\lambda, \phi, \eta_1) = \sum_{\substack{e \in \mathcal{E} \\  s \in \srcset}} y^{(\tau)}_{1, s, e} \log q_1(\delta_e; \lambda k(x_{i_e}, x_{s}; \phi), \eta_1)\;.
\end{align}
The joint optimization in these parameters is 
\begin{align}
    \bigl(\hat{\lambda}^{(\tau + 1)}, \hat{\phi}^{(\tau + 1)}, \hat{\eta}_1^{(\tau + 1)}\bigr) = \argmax_{\lambda, \phi, \eta_1} Q_3^{(\tau)}(\lambda, \phi, \eta_1) \;. \label{eq:em-q3}
\end{align}
The location $\lambda k$ is a product of $\lambda$ and the kernel, so we do not maximize $Q_3^{(\tau)}$ jointly as standard EM would. 
Instead, as GEM allows, each M-step performs one sweep of block coordinate ascent, updating $\phi$, then $\lambda$, then $\eta_1$, each while holding the others at their latest values \cite{meng1993maximum}. The $\lambda$ and $\eta_1$ updates are exact conditional maximizers, so they cannot decrease $Q_3^{(\tau)}$; whether the $\phi$ update does depends on the kernel (\cref{subsubsec:si_em_phi}). 

Write $k_{s,e} \equiv k(x_{i_e}, x_s; \phi)$. 
For the Gaussian noise model 
\begin{align}
    q_1(\delta; \lambda k, \eta) = \frac{1}{\eta \sqrt{2\pi}} \exp \left[ - \frac{1}{2} \left(\frac{\delta - \lambda k}{\eta}\right)^2 \right]\;,
\end{align}
we obtain 
\begin{align}
    Q_3^{(\tau)}(\lambda, \phi, \eta_1) = \sum_{\substack{e \in \mathcal{E} \\  s \in \srcset}} y^{(\tau)}_{1,s,e} \left[-\frac{1}{2\eta_1^2}\bigl(\delta_e - \lambda k_{s,e}\bigr)^2 - \log \eta_1 \right] + C\;. \label{eq:em-q3-gauss}
\end{align}

The updates are as follows: 
\begin{enumerate}
    \item $\hat{\phi}^{(\tau+1)}$. Holding $\lambda$ at $\hat{\lambda}^{(\tau)}$, 
    \begin{align}
        \hat{\phi}^{(\tau + 1)} = \argmax_{\phi} Q_3^{(\tau)}\bigl(\hat{\lambda}^{(\tau)}, \phi, \hat{\eta}_1^{(\tau)}\bigr)\;. \label{eq:em-phi}
    \end{align}
    Under Gaussian noise this is a nonlinear weighted least-squares problem in $\phi$, and it does not involve $\eta_1$. 
    It has no closed form in general, and we solve it with a kernel-dependent method described in \cref{subsubsec:si_em_phi}. 
    \item $\hat{\lambda}^{(\tau+1)}$. Holding the kernel fixed at $\hat{\phi}^{(\tau+1)}$, the location is linear in $\lambda$, so by analogy with \cref{eq:em-mu0-gauss} this is a weighted least-squares problem with solution 
    \begin{align}
        \hat{\lambda}^{(\tau+1)} = \frac{\sum_{e,s} y^{(\tau)}_{1,s,e}\, k_{s,e}\, \delta_e}{\sum_{e,s} y^{(\tau)}_{1,s,e}\, k_{s,e}^2}\;. \label{eq:em-lambda-gauss}
    \end{align}
    \item $\hat{\eta}_1^{(\tau+1)}$. As for the retention arm in \cref{eq:em-eta0-gauss}, the scale is the weighted standard deviation of the residuals, now weighted by $y^{(\tau)}_{1,s,e}$ instead of $y^{(\tau)}_{0,s,e}$: 
    \begin{align}
        \hat{\eta}_1^{(\tau+1)} = \sqrt{\frac{\sum_{e,s} y^{(\tau)}_{1,s,e} \bigl(\delta_e - \hat{\lambda}^{(\tau+1)} k_{s,e}\bigr)^2}{\sum_{e,s} y^{(\tau)}_{1,s,e}}}\;. \label{eq:em-eta1-gauss}
    \end{align}
\end{enumerate}
Since $\hat{\lambda}^{(\tau+1)}$ does not depend on $\eta_1$, steps 2 and 3 give the exact maximizer over $(\lambda, \eta_1)$ given $\hat{\phi}^{(\tau+1)}$. 

For Laplace noise the updates are the same \emph{mutatis mutandis}: just as in \cref{eq:em-mu0-lap,eq:em-eta0-lap}, the weighted squared-error objective is replaced by a weighted absolute-deviation objective. 
Step 1 becomes a nonlinear weighted $L_1$ problem in $\phi$. 
In step 2, $\sum_{e,s} y^{(\tau)}_{1,s,e} |\delta_e - \lambda k_{s,e}| = \sum_{e,s} y^{(\tau)}_{1,s,e} |k_{s,e}|\, |\delta_e / k_{s,e} - \lambda|$, so $\hat{\lambda}^{(\tau+1)}$ is a weighted median of the ratios $\delta_e / k_{s,e}$, with weights $y^{(\tau)}_{1,s,e}|k_{s,e}|$ rather than $y^{(\tau)}_{1,s,e}$. 
Step 3 is the weighted mean absolute deviation: 
\begin{align}
    \hat{\lambda}^{(\tau+1)} &= \operatorname*{weighted\,median}_{(e,s):\,k_{s,e} \neq 0}\Bigl(\bigl\{\delta_e / k_{s,e}\bigr\}, \bigl\{y^{(\tau)}_{1,s,e} |k_{s,e}|\bigr\}\Bigr)\;, \label{eq:em-lambda-lap}\\
    \hat{\eta}_1^{(\tau+1)} &= \frac{\sum_{e,s} y^{(\tau)}_{1,s,e} \bigl|\delta_e - \hat{\lambda}^{(\tau+1)} k_{s,e}\bigr|}{\sum_{e,s} y^{(\tau)}_{1,s,e}}\;. \label{eq:em-eta1-lap}
\end{align}

If $\pi=1$, the retention arm is absent. If $\pi=0$, the influence arm is absent. If $\lambda=0$, the influence arm is centered at zero, but the two noise components can still form a mixture.

\subsubsection{Kernel parameter optimizers}
\label{subsubsec:si_em_phi}

The solution of \cref{eq:em-phi} depends on the kernel.
\begin{itemize}
    \item \textbf{DeGroot} has no kernel parameters, so this step is empty and only $\hat{\lambda}$ is updated, by \cref{eq:em-lambda-gauss} or \cref{eq:em-lambda-lap}.
    \item \textbf{Bounded confidence}: the gain is a step function of the discrepancy with zero gradient almost everywhere, so we maximize over its threshold by golden-section search \cite{kiefer1953sequential} on a width spanning the observed opinion range, with tolerance $10^{-6}$ and at most 500 iterations.
    \item \textbf{SAR and SBC}: We use gradient descent with Adagrad steps on $-Q_3^{(\tau)}$ with learning rate $0.01$ for at most 20 (SAR) or 100 (SBC) steps per EM step, with the optimizer state reset at every EM step and gradient norms clipped to $\sqrt{n_\phi}$, where $n_\phi$ is the number of kernel parameters. The M-step stops after at least 5 steps, once the relative change in the objective is below $10^{-5}$ and the gradient norm is below $0.1$ for 3 consecutive steps.
    \item \textbf{RFF}: For Gaussian noise we solve the responsibility-weighted least-squares problem with a ridge penalty in closed form. For Laplace noise we use the same weighted least-squares step as a surrogate (\cref{SI:model_nonparam_ker}). 
\end{itemize}

\subsection{Model selection}
\label{subsec:si_model_selection}

For out-of-sample model selection, we employ a node-clustered 5-fold cross-validation procedure. A naive split by event would put events from one user's trajectory in both the training and the held-out folds, which leaks information between them. To prevent this, we cluster by node: The data is split into 5 folds. Each node or target within the dataset is randomly assigned to one of the folds, producing 5 folds of approximately equal size. In the case of single-source exposure data, this eliminates any possible source of contamination; in the case of multi-source exposure data, these correlations, weak between nodes, may still exist. 

Each model is trained on 4 of the folds and evaluated on each held-out fold. 

Since the likelihood landscape is in general still non-convex, we employ multiple random restarts for each run. Since the scale of the data differs for each dataset, the parameter initializations are chosen in a range per dataset. Each parametric kernel gets 4 initial values of $\phi$. Since all kernels share a parameter related to the width of the kernel, $\phi_0$ is set to the 25th, 50th, 75th and 90th percentiles of the discrepancies $|d_e|$. For the SBC kernel, which possesses a separate $\phi_1$ to control the steepness, we pair these with $\phi_1 \in \{2, 8, 20, 40\}$. 

For nonparametric models, the ridge penalty is fixed at $10^{-3}$, but the bandwidth $\gamma$ is swept over in a dataset-dependent range which is set to $\{1/4,1/2,1,2,4\} / \mathrm{median}\,|d_e|$. Because RFF relies on random feature sampling, four random seeds are used to draw $R=256$ random features. To prevent overfitting, the lowest $\gamma$ within one standard error of the maximum is selected as the winner \cite{hastie2009elements}. DeGroot and the random-walk null use one initialization each. 

Within each fold, the restart with the highest training MLL per transition is returned as the fold winner; the held-out fold is used only to score it. 

Sweeps are conducted over the model grid
\begin{align*}
\{\text{RFF},\text{SBC},\text{BC},\text{SAR},\text{DeGroot},\text{null}\}\times\{\text{Laplace},\text{Gaussian}\}\times\{\text{no retention},\text{retention}\}.
\end{align*}
For multiple source exposure datasets, the grid is further crossed with
\begin{align*}
\{\text{neighbor aggregation},\text{latent neighbor selection}\}.
\end{align*}

\subsection{EM training}
\label{subsec:si_em_training}

\textbf{EM Convergence Conditions}: Each EM run stops when it reaches a cap on the number of EM steps (600 for the networked panels, 300 for the dyadic ones) or when the relative change in the marginal log-likelihood stays below $10^{-6}$ for 10 consecutive steps.

The kernel-parameter step inside each M-step is described in \cref{subsubsec:si_em_phi}.

\section{More detail on synthetic data experiments}
\label{sec:si_synthetic}
In this section, we show results on kernel inference with synthetic data. To check that the model recovers the kernel and selects the right mechanisms, we fit synthetic data generated from the same data-generating process used for the inverse problem. 
In the networked experiments, we simulate our model on the Les Mis\'erables character interaction network with $N=77$ nodes for $T=500$ time steps. In the dyadic experiments, each update has a single source.

\subsection{Kernel-family selection}
\label{app:synthetic_selection}

\Cref{tab:si_synthetic_2e} fits every kernel, arm and noise model to one networked synthetic panel. \Cref{tab:si_synthetic_discrimination} repeats the selection across true kernels, sample sizes and influence rates $\pi$.

\begin{table}[H]
  \centering
  {\small\setlength{\tabcolsep}{3pt}
\begin{tabular}{lrrrrr}
\toprule
arm / noise & DeGroot & BC & SBC & SAR & RFF \\
\midrule
mean field / Gaussian & -0.563 & -0.561 & -0.548 & -0.548 & -0.548 \\
mean field / Gaussian mixture & -0.288 & -0.286 & -0.273 & -0.273 & -0.273 \\
mean field / Laplace & -0.518 & -0.524 & -0.517 & -0.557 & -0.523 \\
mean field / Laplace mixture & -0.319 & -0.315 & -0.294 & -0.459 & -0.294 \\
pairwise / Gaussian & -0.547 & -0.498 & -0.490 & -0.527 & -0.427 \\
pairwise / Gaussian mixture & -0.234 & -0.186 & -0.038 & -0.069 & -0.039 \\
pairwise / Laplace & -0.525 & -0.475 & -0.505 & -0.525 & -0.379 \\
pairwise / Laplace mixture & -0.205 & -0.158 & \textbf{0} & -0.044 & \textbf{0} \\
\bottomrule
\end{tabular}}
\par\smallskip{\footnotesize Synthetic data, truth: SBC kernel, Laplace mixture noise, $\pi=0.5$, $\lambda=0.5$, $\eta_1=0.01$. Gap in held-out MLL per transition to the best cell (bold, 0). }

  \caption{Held-out MLL per transition on a synthetic networked panel, for every kernel (columns) and every arm and noise model (rows), as a gap to the best cell. The data come from a sigmoidal bounded-confidence kernel with latent selection and latent retention under Laplace noise. Mean field is the neighbor-aggregation arm and pairwise the latent-selection arm; the mixture noise models include latent retention.}
  \label{tab:si_synthetic_2e}
\end{table}

\begin{table}[H]
  \centering
  {\small\setlength{\tabcolsep}{5pt}
\begin{tabular}{lccc}
\toprule
setting & true DeGroot & true SBC & true SAR \\ & \multicolumn{3}{c}{selected kernel / noise recovered} \\
\midrule
network, $\pi{=}$0.5 & BC $\times$ / \checkmark & SBC \checkmark / \checkmark & SAR \checkmark / \checkmark \\
network, $\pi{=}$1 & SBC $\times$ / \checkmark & SBC \checkmark / \checkmark & SAR \checkmark / \checkmark \\
network, $\pi{=}$0.5, SAR side study: mean-field arm & --- & --- & RFF $\times$ / \checkmark \\
network, $\pi{=}$0.5, SAR side study: pairwise RFF & --- & --- & RFF $\times$ / \checkmark \\
dyadic, $n{=}10^3$, $\pi{=}$0.5 & DeGroot \checkmark / \checkmark & SBC \checkmark / \checkmark & SAR \checkmark / \checkmark \\
dyadic, $n{=}4{\times}10^3$, $\pi{=}$0.5 & DeGroot \checkmark / \checkmark & SBC \checkmark / \checkmark & SAR \checkmark / \checkmark \\
dyadic, $n{=}10^5$, $\pi{=}$0.5 & SAR $\times$ / \checkmark & SBC \checkmark / \checkmark & SAR \checkmark / \checkmark \\
dyadic, $n{=}10^3$, $\pi{=}$1 & DeGroot \checkmark / $\times$ & SBC \checkmark / $\times$ & SAR \checkmark / \checkmark \\
dyadic, $n{=}4{\times}10^3$, $\pi{=}$1 & SAR $\times$ / \checkmark & SBC \checkmark / \checkmark & SAR \checkmark / $\times$ \\
dyadic, $n{=}10^5$, $\pi{=}$1 & SAR $\times$ / \checkmark & SBC \checkmark / \checkmark & SAR \checkmark / \checkmark \\
\bottomrule
\end{tabular}}

  \caption{Model selection on one synthetic dataset per setting. Each cell gives the kernel with the highest held-out MLL when the column's kernel generated the data, whether it is the true kernel, and whether the selected noise model is the true one. The two side-study rows restrict the fit to one arm or to the \rff{} kernel. With one dataset per setting, DeGroot truth is not always recovered: in five of the eight settings with a DeGroot truth, another kernel is selected instead. In all five the selected kernel beats DeGroot by less than $10^{-4}$ nats per transition: its fit is effectively linear over the data (\sar{}, for instance, with a fitted width near 13).}
  \label{tab:si_synthetic_discrimination}
\end{table}

\subsection{Nonparametric recovery}

\begin{figure}[t]
    \centering
    \includegraphics[width=\linewidth]{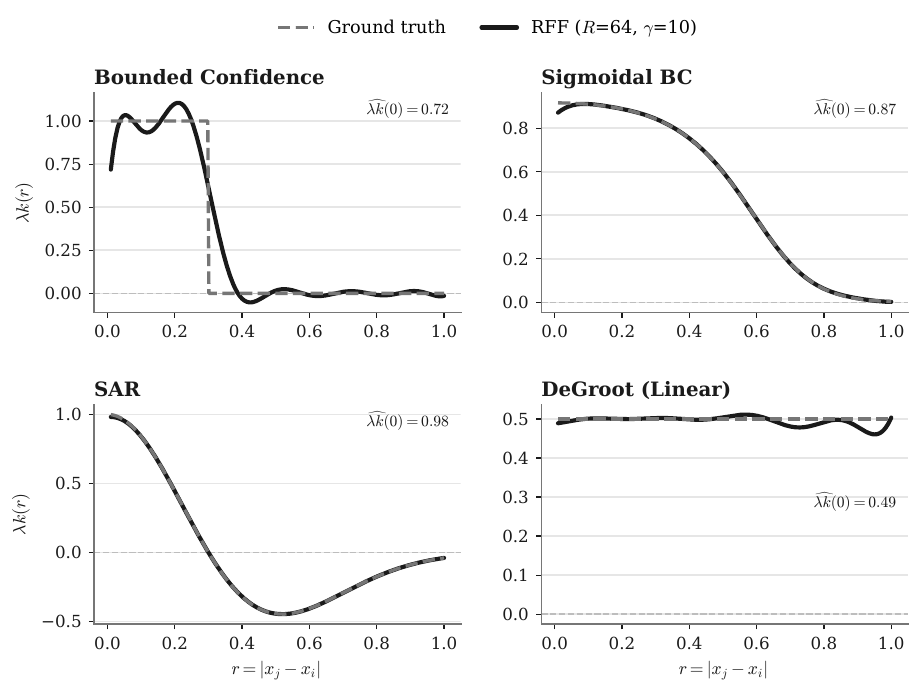}
    \caption{Recovery of planted kernels on simulated data with the nonparametric random Fourier feature (\rff{}) kernel, using the one-dimensional gain representation. Dashed: the true kernel. Solid: the \rff{} fit. Curves show the influence $\lambda k(r)$, since only the product is identified; the simulations have no separate scale, so the dashed curve is the planted kernel itself. The corner text gives the fitted $\lambda k(0)$. These fits use $R = 64$ features and $\gamma = 10$, not the $R = 256$ of the empirical fits.}
    \label{fig:si_nonparametric_recovery}
\end{figure}

Figure~\ref{fig:si_nonparametric_recovery} shows recovery on simulated data using the one-dimensional gain representation. The \rff{} kernel recovers every planted kernel except bounded confidence well. The sharp cutoff of bounded confidence is a high-frequency feature that a smooth basis struggles to approximate, and the \rff{} fit rings near the step. A wavelet basis could capture sharp kernels better.

\subsection{Omitting a latent mechanism shrinks the kernel}
\label{app:omitting_mechansims}
\paragraph{Omitting retention shrinks social influence}
A regression fit by mean squared error estimates the conditional mean response. Marginalizing \eqref{eq:si_emission} over $z_e$ gives:

\[
\mathbb{E}[\upd \mid \pull] = (1-\pi) \mu_0 + \pi \lambda \gain\,\pull.
\]

When $\mu_0 \approx 0$,

\[
\mathbb{E}[\upd \mid \pull] \approx \pi \lambda \gain\,\pull.
\]

A model that omits retention will attenuate the susceptibility by a factor of $\pi$:

\[
\hat{\lambda} = \pi \lambda.
\]

Figure~\ref{fig:synthetic_misspecification} shows the effect: omitting latent retention produces a lower-magnitude kernel. The shape of the kernel is close to the truth. The smaller kernel is harder to tell apart from a linear one.

\paragraph{Omitting source selection shrinks social influence}

When the true interaction is with a single source $\src$, the true discrepancy is

\[
\pull[\src] = \glyphOpinion_{\src} - \glyphOpinion_{\glyphTarget}.
\]

An aggregate model instead uses

\[
\bar{\glyphPull} = \bar{\glyphOpinion} - \glyphOpinion_{\glyphTarget},
\qquad
\bar{\glyphOpinion}
  = \frac{1}{|\partial \glyphTarget|}\sum_{\src \in \partial \glyphTarget} \glyphOpinion_{\src},
\]

where $\bar{\glyphOpinion}$ is the mean opinion over the sources available to $\glyphTarget$. Thus, $\pull[\src] = \bar{\glyphPull} + \dev$, where $\dev$ is the deviation of the selected source from that mean.

If the true response is $f(\glyphPull)=\lambda\,\kernel(|\glyphPull|)\,\glyphPull$, the regression on $\bar{\glyphPull}$ is

\[
\bar{f}(\glyphPull) = \mathbb{E}_{\glyphDev}\bigl[f(\bar{\glyphPull} + \glyphDev)\bigr].
\]

When $\glyphDev$ is independent of $\bar{\glyphPull}$, this is a convolution \cite{Delaigle2006}:

\[
\bar{f} = f \star p_{\glyphDev}.
\]

Averaging over $\glyphDev$ cannot add variation, so the total variation of $\bar{f}$ is at most that of $f$:

\[
\text{TV}(\bar{f}) \leq \text{TV}(f).
\]

\Cref{fig:synthetic_misspecification}b shows this: aggregating over the available sources cuts the size and the total variation of the inferred kernel by more than half, while modeling source selection returns a nearly identical kernel.
\section{Dataset preparation}
\label{app:dataset_prep}

\subsection{Overview}

The most important distinction among the datasets concerns whether the opinion dynamics are synchronous or asynchronous. In some settings, such as Congressional data \citep{NokkenPoole2004} or controlled numerical estimation experiments \citep{adams2022person,Becker2017,Becker2019}, opinion values are reported for all nodes at regular intervals. In such cases, the dynamics are naturally modeled as synchronous updates, with each reporting period marking a single timestep in the model. Under synchronous dynamics, all nodes update their states at every timestep.

Other datasets, particularly those generated from social media, do not neatly fall into a synchronous-update paradigm. Instead, users express their opinions through posts, and each post may represent an updated opinion. In this case, each post can be considered a single timestep, and users can be thought of as updating their opinions asynchronously. Reddit datasets such as CMV and SPINOS \cite{Sakketou2022} represent asynchronous dynamics.

The selected datasets also quantify opinions in radically different ways. Numerical estimation tasks, such as those used in the wisdom-of-crowds experiments \citep{Becker2017,Becker2019}, follow a long tradition of numerical estimation experiments in social psychology and sociology. Numerical estimation provides a directly quantifiable measure of a participant's opinion that does not require additional inference or derived proxy metrics. However, there are some drawbacks. First, the values in numerical estimation tasks are unbounded, taking arbitrary real or rational values. This is problematic because the scale and support of the opinion kernel are not standardized. In political tasks, the numerical value also does not cleanly map onto partisan alignment.

The second source of data, also widely used in the social psychology literature, consists of ordinal categorical measures of opinion. Such scales are clearer, easier to self-report, and have a direct interpretation. However, they do not lend themselves naturally to most models of interaction kernels in opinion dynamics, which usually assume real-valued opinions.

In many settings, especially on social media, opinions are expressed through text. In this case, there is no direct numerical quantification of opinion in the data, and an indirect method must be used to convert a text post into an opinion value. Several approaches exist. For example, \citet{Monti2020} treat opinions as latent variables and treat interactions, such as posting in a particular subreddit, as observable actions driven by latent opinion. \citet{Kozitsin2021} use a VK dataset and constructs an opinion network using certain partisan accounts as anchors to project each user into a one-dimensional opinion space.

One of the most promising directions for opinion quantification uses tools developed in natural language processing for stance detection. Stance detection classifies the position of a piece of text relative to a specific anchor or reference position \cite{ALDayel2021}.

Dataset sizes are listed in \Cref{tab:datasets}.

\subsection{General dataset preparation and filtering}

\begin{table}[htbp]
\centering
\footnotesize
\setlength{\tabcolsep}{3pt}
\renewcommand{\arraystretch}{0.92}
\caption{Transitions per panel: active transitions (target observed at both ends of the step), the share of those removed as exact zeros, and the events kept after that removal. Held-out scored counts are smaller on \house{}, \cmv{}, \spinos{} and \markets{}, where events with an empty candidate set carry no likelihood term.}
\label{tab:si_panel_summary}
\begin{tabular}{@{}lrrr@{}}
\toprule
Panel & Active & \shortstack{Removed\\(\%)} & \shortstack{Events\\kept} \\
\midrule
\adamsLo{}    & 1,051  & 0.0  & 1,051 \\
\adamsMid{}   & 1,097  & 0.0  & 1,097 \\
\adamsHi{}    & 1,114  & 0.0  & 1,114 \\
\takacsCtl{}  & 1,140  & 60.0 & 456 \\
\takacsDis{}  & 1,410  & 61.4 & 544 \\
\takacsInt{}  & 840    & 61.0 & 328 \\
\takacsOne{}  & 1,204  & 56.5 & 524 \\
\becker{}     & 4,296  & 30.3 & 2,996 \\
\vkg{}        & 16,121 & 41.1 & 9,499 \\
\vkc{}        & 18,689 & 32.4 & 12,627 \\
\senate{}     & 987    & 0.5  & 982 \\
\house{}      & 4,040  & 0.7  & 4,010 \\
\cmv{}        & 2,806  & 16.3 & 2,348 \\
\spinos{}     & 1,006  & 45.8 & 545 \\
\kozitsin{}   & 6,000  & 15.6 & 5,067 \\
\markets{}    & 18,186 & 21.5 & 14,277 \\
\bottomrule
\end{tabular}
\end{table}

In the case of networked temporal datasets, the candidate set of exposures is unclear. For each, we construct a network; details are recorded in the respective sections. We consider transitions from $x_e^-$ to $x_e^+$. We consider all events where the node $i$ has posted at least once before to provide a value of $x_e^-$. The value of a current event is extracted from the post or comment (details are specific to each dataset). The set of candidate exposures consists of all of those neighbors of node $i$ with at least one post with an opinion value that was made before event $e$.

Exact-zero updates ($\delta_e = 0$) form an atom at zero. We remove them in preprocessing rather than modeling them in the likelihood, so the model can be read as having a third option, an exact stay, that we handle outside the fit. Every fitted quantity is therefore conditional on a nonzero update. In particular $\hat\pi$ is the share of influence among nonzero updates; with a share $z$ of exact zeros, the share of all updates that are influence is at most $(1-z)\hat\pi$. The share of exact zeros removed from each panel is given in Table~\ref{tab:si_panel_summary}. The share is taken over the transitions that would otherwise be scored: a node that holds an observed opinion at both ends of the step.

Some panels record opinions on a grid: integer or ordinal responses in the Tak\'acs, Vande Kerckhove and SPINOS data, and Nokken--Poole scores that the source rounds to three decimals in the Senate and House data. A continuous density on gridded data can grow without bound, so on these panels we \emph{de-quantize} \cite{Ho2019}: we add noise drawn uniformly from $[-h/2, h/2]$, where $h$ is the grid step. The grid steps are $h=10^{-3}$ for \senate{} and \house{}, $h=0.01$ for the Tak\'acs panels and \vkg{}, $h=0.002$ for \vkc{} and $h=0.25$ for \spinos{}. By Sheppard's correction the added noise inflates the variance of each opinion by $h^2/12$, and so the variance of an update, a difference of two opinions, by $h^2/6$. This is negligible on every panel except \spinos{}. The \adams{}, \becker{}, \kozitsin{}, \markets{} and \cmv{} panels are continuous and are not de-quantized. 

A brief description of the nature of each dataset and our cleaning methods is outlined below:

\subsection{\texorpdfstring{\adams{}}{ADAMS}: Adams et al.\ (2022)\texorpdfstring{~\cite{adams2022person}}{}}
\label{app:data_adams}

\subsubsection{Description}
We test our kernel on several empirical datasets. The first was collected in a controlled experiment designed to test opinion dynamics models~\cite{adams2022person}. In this study, participants were asked to locate a hidden dot within a box. Participants first saw a blue circle representing an 80\% confidence interval that was always reliable. They then saw a red circle, also purporting to be an 80\% confidence interval, but with variable reliability signaled by a traffic light: green indicated 80\% reliability, yellow 50\%, and red 20\%.
Participants then drew their own circle indicating where they believed the dot to be located. Each participant's final placement is treated as their opinion, encoded as the center coordinate of the circle they drew. Interactions in this dataset are purely dyadic, and there is no fixed network structure, making this the simplest and cleanest test case for our kernel inference framework.

\subsubsection{Preparation}

We treat the initial opinion as the center of the blue circle. The discrepancy $\pull$ is the distance between the center of the blue and red circles, while the update $\delta_e$ is the distance in pixels between the center of the initial blue circle and the user's updated circle guess. Each level of confidence is analyzed as its own dataset: \adamsLo{} (red-circle reliability 20\%, 1,051 trials), \adamsMid{} (50\%, 1,097 trials) and \adamsHi{} (80\%, 1,114 trials), 3,262 trials in all. The remaining 498 trials of the 3,760 in the original data were run at other reliability levels and are not used. We normalize by the largest observed discrepancy, 350.33 pixels, to ensure all distances are within $[0,1]$. 

\subsection{\texorpdfstring{\becker{}}{BECKER}: Becker et al.\ (2017)\texorpdfstring{~\cite{Becker2017}}{}}
\label{app:data_becker}

The next dataset comes from controlled laboratory studies on the ``wisdom of crowds.'' In \citet{Becker2017}, participants made numerical estimates (e.g., population sizes, geographic facts) and then observed estimates from sources in either decentralized (random regular) or centralized (hub-and-spoke) network structures. While the random regular topologies were not recoverable from available data, the hub-and-spoke networks were. We extract only the spoke nodes and treat them as a dataset, along with a single-source exposure where each spoke node is treated independently. There are only 59 hub nodes (one per trial), too few to analyze the hubs on their own in a multi-source exposure setting. The hub stays in the network as a source, but its own transitions are masked, so every scored update is a spoke responding to its one neighbor, the hub. Following the methods of the original paper, we z-score responses per question to normalize scale. Participants whose per-round standard deviation exceeds 3 are dropped (67 of 2,274).
Some respondents failed to complete every question every round. Respondents are excluded from analysis when no data exists.

\subsection{\texorpdfstring{\takacs{}}{TAKACS}: Tak\'acs et al.\ (2016)\texorpdfstring{~\cite{Takacs2016}}{}}

The study by Tak\'acs et al.~\cite{Takacs2016} analyzes the effect of social influence in a controlled setting to detect the presence of repulsive interactions. The data consist of two studies conducted with sociology undergraduates at the University of Groningen. In both studies, participants are asked to give their opinion on questions with no objective true answer on 83 different issues and asked to rate their agreement on a scale from 0 to 100. They are then exposed to a stimulus and solicited to give a revised estimate. In the first study, \takacsOne{}, 108 students were given a web questionnaire. Participants are told the stimulus comes from another participant but it is in fact a pre-recorded source. The second study pairs real participants together through chat software. Each participant is exposed to the other's opinions and asked to update their own in line with the first experiment. Study 2 uses a within-subject disliking treatment. The control and the treatment alternate rounds. \takacsCtl{} follows the same pattern as the first study. \takacsDis{} applies a disliking treatment which consists of three components: participants are told their partner is in a different major, a round of the Prisoner's Dilemma and the choice to send a stigmatizing message. \takacsInt{} drops rounds 1 and 9 (the first and last rounds) so the control and disliking arms have the same number of rounds. Because the treatment alternates, it is an exact function of round parity: disliking runs in the odd rounds and control in the even ones. A difference between the two arms can therefore also reflect a difference between odd and even rounds. We divide opinions by 100 to normalize their range to $[0, 1]$ and then de-quantize them. 

\subsection{\texorpdfstring{\vkg{}}{VK-gauging} and \texorpdfstring{\vkc{}}{VK-counting}: Vande Kerckhove et al.\ (2016)\texorpdfstring{~\cite{VandeKerckhove2016}}{}}

In Vande Kerckhove et al.~\cite{VandeKerckhove2016} participants are grouped in sets of 6 and asked to perform two types of numerical estimation experiments. The experiment consists of 30 games per group, each game consisting of three rounds. In \vkg{}, participants are shown a picture consisting of patches of different colors and asked to estimate the percentage of the picture which contains a particular color. They are then shown all the estimates from the other 5 participants in their group and asked to revise their answer. We treat each participant group as a 6-node clique. Opinion variables are the percentage of the image, between 0 and 100. In \vkc{} participants are shown an image consisting of discrete objects and asked to estimate the total number of objects. Opinions are raw counts of objects between 0 and 500. The authors use a trustworthiness filter based on correlation with the truth to remove participants acting in bad faith, and remove participants who completed less than half of their assigned games. We do not replicate this filter.

The data cover 64 groups (1,920 games) for gauging and 71 groups (2,130 games) for counting, from the S1 Dataset released with the 2020 correction to the paper.

We scale gauging responses by 100 and counting responses by 500, so opinions lie in $[0,1]$, and we treat each game as a complete graph on the six players. Each player has two transitions per game. Responses are integers, so we de-quantize with $h=0.01$ for gauging and $h=0.002$ for counting. Participants who did not respond in some round are recorded as unobserved (26.2\% of gauging and 23.4\% of counting cells) and are never scored. Of the 16,121 active gauging transitions, 9,499 remain after removing exact zeros, and of the 18,689 active counting transitions, 12,627 remain.

\subsection{\texorpdfstring{\kozitsin{}}{VKontakte}: Kozitsin (2021)\texorpdfstring{~\cite{Kozitsin2021}}{}}
\label{app:data_vk}

The \kozitsin{} data come from \citet{Kozitsin2021}. Each user has a score in $[0,1]$ at three waves in 2018 (February, July and December), estimated by a supervised model from the user's subscriptions to information sources rather than reported in a survey, together with the user's friendship network. For each selected user we build a 1-hop ego graph consisting of their immediate neighbors. We select nodes with a degree capped at 100 for computational efficiency. The friendship graph is a single frozen snapshot of the VK social network and is constant across steps. Only the ego's transitions are scored, so each ego contributes two events; the neighbors serve as candidate sources but are not scored themselves. Our dataset consists of 3,000 randomly sampled ego networks containing 6,000 ego transitions, of which 5,067 remain after removing exact zeros. Unlike the original study, we do not apply its remarkable-shift or common-stimulus filters, because we want to quantify these as part of our model. 

\subsection{\texorpdfstring{\senate{}}{SENATE} and \texorpdfstring{\house{}}{HOUSE}: Nokken--Poole scores\texorpdfstring{~\cite{NokkenPoole2004}}{}}
\label{app:data_congress}

Like \kozitsin{}, the remaining datasets come from uncontrolled, observational settings. Nokken-Poole first dimension scores~\citep{NokkenPoole2004} run from $-1$ (liberal) to $1$ (conservative). A statistical model infers them from roll-call votes, giving each member of Congress one continuous latent opinion per term. We assign one observation per legislator per congressional term, yielding $T=12$ time steps for each of the Senate and the House, one per Congress from the 108th through the 119th. The 119th Congress is still in session, so its Nokken--Poole scores are provisional and will change as more roll calls are recorded.

For network structure, we use the bill-cosponsorship network. Sponsor and cosponsor lists come from the GovInfo bulk bill-status data (\texttt{govinfo.gov/bulkdata/BILLSTATUS}) for the 108th through 119th Congresses, and legislator identifiers are matched to Voteview ICPSR numbers through the \path{unitedstates/congress-legislators} and Voteview member files. A weighted network is constructed where the weight of a pair of legislators is the number of bills that one sponsored and the other cosponsored, summed over the 12 Congresses, and an edge exists when this count is 3 or more. We then apply a disparity filter~\cite{Serrano2009} to sparsify the graph with $\alpha=0.1$. Congress is an open system: 
legislators enter and leave Congress over time. We include each legislator only for the duration of their terms. Our methods overlap with those of \cite{Liu2015}, who perform a similar analysis although they use two networks: committee co-membership and physical proximity of seating in the Senate chamber. 

\subsection{\texorpdfstring{\markets{}}{MARKETS}: Manifold prediction markets}
\label{app:data_manifold}

The prediction-market data from Manifold are similar to the synchronous settings above in that they provide repeated, time-stamped measurements of beliefs. Markets ask users to buy shares in the outcome of an event, and users receive payouts according to whether the event occurs. We use each trade as a noisy report of the trader's belief that the event will occur.

On Manifold, users purchase shares in propositions (e.g., ``Will AI generate a feature-length film by 2028?''). Each market specifies resolution conditions and a timeframe. Users buy YES or NO shares at prices reflecting the current probability estimate; shares pay out at face value if correct and zero otherwise. Unlike real-money platforms such as Polymarket, Manifold uses a play-money currency called Mana, which cannot be exchanged for cash but can be used to donate to charity.

All data were collected via the Manifold API in April 2026 (14 markets on 4 April and 62 on 10 April). Candidate markets were resolved binary markets with a constant-product (CPMM) market maker, at least 30 unique bettors, at least 14 days since creation and a duration of at least 7 days. Markets on sports, political figures, finance and news-driven topics were excluded by question keywords and group tags. Among the candidates we kept those with at least 5 commenting users and at least 50 usable bets. We keep only users with at least 5 bets. After requiring $T\ge 50$ and $N\ge 5$ in the inference data, 76 markets remain, with $N$ up to 99 users and $T$ between 50 and 2,591 bets per market.

\subsubsection{Implied probability of a trade}

Each trade gives one observation of the trader's belief. We take the trade's implied probability of YES, the amount paid divided by the shares received, and use one minus that value for NO shares. This is the average fill price of the trade, not the market probability after it. We clip it to $[0,1]$. A user's first trade sets their initial opinion and is not scored; every later trade is a scored transition.

The network is the reply network: we place an edge between users $i$ and $j$ if one replies to a comment of the other. We drop edge weights and direction, and use no disparity filter. Only users who both trade and appear in a reply edge are nodes.

\subsection{\texorpdfstring{\spinos{}}{SPINOS}: Sakketou et al.\ (2022)\texorpdfstring{~\cite{Sakketou2022}}{}}
\label{app:data_spinos}

We next consider data collected from Reddit. We use the Subtle Polarity and Intensity Opinion Shifts (SPINOS) dataset \cite{Sakketou2022}. The SPINOS data contains stance annotations collected from human annotators. The dataset contains 3,526 annotated posts from 638 users, collected from subreddits on 8 topics. The stance of commenters on selected subreddits was evaluated by human annotators on a five-point scale (strongly against, against, undecided, in favor, strongly in favor), plus a label for posts whose stance could not be inferred. The scale is mapped uniformly to $\{0, 0.25, 0.5, 0.75, 1\}$. Posts whose stance cannot be inferred are dropped. Undecided posts and posts most annotators mark as sarcastic carry no stance: they create no transition, and the user's opinion carries forward unchanged. Stance-bearing opinions therefore take the values $\{0, 0.25, 0.75, 1\}$. SPINOS records the parent of each comment, and we connect two users when one replies to the other. The SPINOS dataset offers several major advantages. First, it offers a long time series per user: in the four retained topics, 391 users write 1,580 posts, 1,347 of them stance-bearing, about 3.4 per user, allowing many transitions to be observed. Second, annotations are made by several human annotators, and the inter-annotator agreement is measured and confirmed to be high, providing a level of reliability not possible with automatic methods of stance detection. For our purposes, the largest limitation of this dataset is the discrete nature of the ratings. The model assumes a continuous density for the update, which does not account for discrete scores; de-quantization only partly addresses this. In the future, a classifier could be trained on the underlying comments, with the SPINOS ratings used as a training dataset. The logits from this classifier would provide a continuous measure of opinion calibrated with highly reliable annotations. We fit the four topics as one panel.

The SPINOS dataset collects data across 8 topics: abortion (1,293 posts), feminism (572), Brexit (467), capitalism (383), climate change (243), guns (226), nuclear energy (184) and veganism (158). We keep the four largest: abortion, feminism, Brexit and capitalism. After dropping posts whose stance annotators could not infer, these four have 864, 258, 218 and 240 usable posts from 187, 77, 67 and 61 users (one user posts in two topics, so these sum to 392), giving time series of length $T=865$, $259$, $219$ and $241$ respectively (the initial state plus one step per post).

Each topic has its own reply network over its users.

\subsection{\texorpdfstring{\cmv{}}{CMV}: r/ChangeMyView stance detection}
\label{app:data_cmv}

We apply stance detection to a final Reddit dataset. We consider posts on the subreddit r/ChangeMyView (CMV). In this subreddit, a user, referred to as the original poster (OP), creates a thread explaining a position they hold, such as ``I think nuclear energy is the best path to decarbonization.'' Commenters then provide arguments intended to change the OP's opinion. If a commenter is successful in changing the OP's opinion, the OP awards them a $\Delta$. The discussion on CMV is particularly convenient for stance detection because the original post provides an anchor. The stance of each comment can be evaluated relative to the text of the OP. Each comment is scored by the \href{https://huggingface.co/alex2awesome/stance-detection-classification-model}{stance-detection classification model} \cite{SpangherStanceModel}. The classifier has four output classes: \textit{Discuss} (asking questions or providing context), \textit{Affirm} (agreeing with the discussed content), \textit{Neutral} (no clear stance), and \textit{Refute} (arguing against the discussed content). For every comment, we assign a continuous score by taking the logits for the Affirm and Refute classes, normalizing them, and computing the opinion as:
\[
\frac{1 + P(\text{Affirm}) - P(\text{Refute})}{2},
\]
which lies in $[0,1]$. Transitions that start or end at exactly 0.5 are masked.

We use the Reddit API to extract posts from $r/CMV$. We collect 18 threads, with 5,958 commenting users in total and between 342 and 2,246 timesteps per thread, and treat the 18 threads as one panel. 

We build one network per thread. An edge exists between two users if they reply in the sub-tree on the same top-level comment.

\section{How the latent mechanisms change the kernel}
\label{app:mechanism_evidence}

\Cref{tab:size_curvature} tracks the size $S$ and curvature $C$ of the \rff{} kernel up the ladder. This section gives the data behind the two readings of it in the main text: retention changes the shape of the kernel, its relative curvature $C/S$, only when retained updates depend on distance, and selection changes the shape only where the neighborhood mean is a poor summary of the neighbors.

\subsection{Retention}
\label{app:retention_shape}

Without a retention arm the influence kernel has to describe retained and influenced updates at once. A Laplace fit finds the conditional median update. At a discrepancy $d$ it sees the median of the retention/influence mixture divided by $d$. If the influence share $\pi$ is the same at every $d$, retention only rescales the size of the kernel and does not change its shape, and the inferred kernel will match this empirical median. \Cref{tab:si_retention_median} compares it with the fit itself on the three \adams{} datasets.

On \adamsLo{} the median matches the fit. Since influenced updates are a minority, the median is dragged down by the retention mode. The kernel without retention is flat with a kink near zero. It is about seven times smaller than it would be otherwise. On \adamsMid{} and \adamsHi{} the constant-$\pi$ median does not match. The influence share that matches the fit falls with distance: on \adamsHi{} from 0.95 near the source to 0.56 at $|d| = 0.6$, so the retention rate rises from 0.05 to 0.44. On these two datasets, agents are more likely to retain their opinion when the stimulus is far away, creating a spurious downward trend in the kernel without retention. 

\begin{table}[ht]
  \centering
  \small
  \begin{tabular}{llccccc}
\toprule
 & & $|d|=0.1$ & $|d|=0.2$ & $|d|=0.3$ & $|d|=0.45$ & $|d|=0.6$ \\
\midrule
\multicolumn{7}{l}{\adamsLo{} ($\hat\pi=0.45$, $\hat\lambda=0.21$)} \\
 & median at constant $\pi$ & 0.05 & 0.04 & 0.03 & 0.03 & 0.02 \\
 & fitted, no retention & 0.03 & 0.03 & 0.03 & 0.03 & 0.03 \\
 & data median $\delta/d$ & 0.03 & 0.04 & 0.04 & 0.03 & 0.02 \\
 & influence share needed &  & 0.36 & 0.44 & 0.48 & 0.49 \\
\addlinespace
\multicolumn{7}{l}{\adamsMid{} ($\hat\pi=0.54$, $\hat\lambda=0.41$)} \\
 & median at constant $\pi$ & 0.12 & 0.11 & 0.11 & 0.12 & 0.16 \\
 & fitted, no retention & 0.23 & 0.14 & 0.11 & 0.11 & 0.11 \\
 & data median $\delta/d$ & 0.30 & 0.26 & 0.11 & 0.15 & 0.07 \\
 & influence share needed & 0.76 & 0.60 & 0.53 & 0.52 & 0.52 \\
\addlinespace
\multicolumn{7}{l}{\adamsHi{} ($\hat\pi=0.81$, $\hat\lambda=0.55$)} \\
 & median at constant $\pi$ & 0.42 & 0.48 & 0.50 & 0.52 & 0.53 \\
 & fitted, no retention & 0.53 & 0.51 & 0.48 & 0.44 & 0.41 \\
 & data median $\delta/d$ & 0.53 & 0.53 & 0.53 & 0.39 & 0.35 \\
 & influence share needed & 0.95 & 0.87 & 0.76 & 0.62 & 0.56 \\
\bottomrule
\end{tabular}

  \caption{Retention on the three \adams{} panels. \emph{Median at constant $\pi$}: the median of the fitted retention/influence mixture, divided by $|d|$. \emph{Fitted, no retention}: the \rff{} gain at the heavy-tail rung, mean over CV folds. \emph{Data median}: the median of $\delta\,\mathrm{sign}(d)/|d|$ in bins centered near each $|d|$. \emph{Influence share needed}: the $\pi(d)$ at which the mixture median equals the fitted gain, with the influence arm held at $\hat\lambda$; blank where none does. $\hat\pi$, $\hat\lambda$ and the noise scales come from the better of the DeGroot and \rff{} fits with retention, on its CV fold with the highest training MLL. \Cref{tab:si_winners_compact} uses the same fold rule but selects over every kernel, so its $\hat\pi$ and $\hat\lambda$ can differ.}
  \label{tab:si_retention_median}
\end{table}


\subsection{Selection}
\label{app:selection_support}

A mean-field kernel sees only the discrepancy to the neighborhood mean, so the shape is only identifiable up to the largest discrepancy. A pairwise kernel sees each neighbor's own discrepancy. When neighbors sit on both sides of a user the mean is pulled in toward the user, and the pairwise kernel is fit at distances the mean-field kernel never reaches. \Cref{tab:si_selection_support} measures this for every networked panel, and compares the size and curvature of the two kernels over the shared support. On the shared range selection increases the size of every kernel on every panel but \cmv{}. \senate{} and \house{} are the panels where the mean is a poor summary: 21\% and 25\% of neighbor distance lies on the far side of the target from the neighborhood mean. Single neighbor discrepancies reach about twice as far as the mean does, and 14\% and 11\% of neighbor weight falls past the mean-field support. On \senate{} the shared-range curvature rises because the kernel's slope reverses. On \house{} the curvature is unchanged, and the bounded-confidence falloff lies past the mean-field support. On \markets{} and the two \vk{} games neighbors sit on the same side as their mean and almost none fall past its support, and selection mostly rescales the kernel; \vkc{} also gains some curvature. \cmv{} has the most weight past the support, but its opinions sit at the extremes of the range: selection removes curvature there. 

\begin{table}[ht]
  \centering
  \small
  \begin{tabular}{lrccccc}
\toprule
panel & degree & past support & opposite side & reach & $S$ & $C$ \\
\midrule
\vkg{} & 4 & 0.02 & 0.00 & 1.2 & 0.55 $\to$ 0.59 & 0.55 $\to$ 0.55 \\
\vkc{} & 4 & 0.03 & 0.02 & 1.2 & 0.27 $\to$ 0.36 & 0.079 $\to$ 0.26 \\
\senate{} & 14 & 0.14 & 0.21 & 1.7 & 0.17 $\to$ 0.23 & 0.25 $\to$ 0.43 \\
\house{} & 21 & 0.11 & 0.25 & 2.1 & 0.14 $\to$ 0.18 & 0.17 $\to$ 0.15 \\
\cmv{} & 63 & 0.26 & 0.00 & 1.3 & 2.1 $\to$ 1.1 & 1.5 $\to$ 0.19 \\
\kozitsin{} & 12 & 0.04 & 0.14 & 1.3 & 0.0022 $\to$ 0.0023 & 0.0020 $\to$ 0.0021 \\
\markets{} & 3 & 0.01 & 0.03 & 1.3 & 0.61 $\to$ 0.74 & 1.2 $\to$ 0.94 \\
\bottomrule
\end{tabular}

  \caption{Neighbor geometry on the networked panels, over updates with more than one neighbor. \emph{Degree}: median number of neighbors. \emph{Past support}: share of neighbor weight at a gap beyond the end of the mean-field kernel's support; each update is weighted by its posterior influence probability and split across neighbors by the prior. \emph{Opposite side}: median share of neighbor distance on the other side of the user from the neighborhood mean. \emph{Reach}: 99th percentile of single-neighbor distance over that of the distance to the mean. \emph{$S$, $C$}: size and curvature of the \rff{} kernel before and after selection, as in \Cref{tab:size_curvature} but over the mean-field kernel's range only.}
  \label{tab:si_selection_support}
\end{table}

\onecolumn\raggedbottom
\section{Full empirical results}
\label{app:empirical_results}

\begin{figure}[H]
  \centering
  \includegraphics[width=\textwidth]{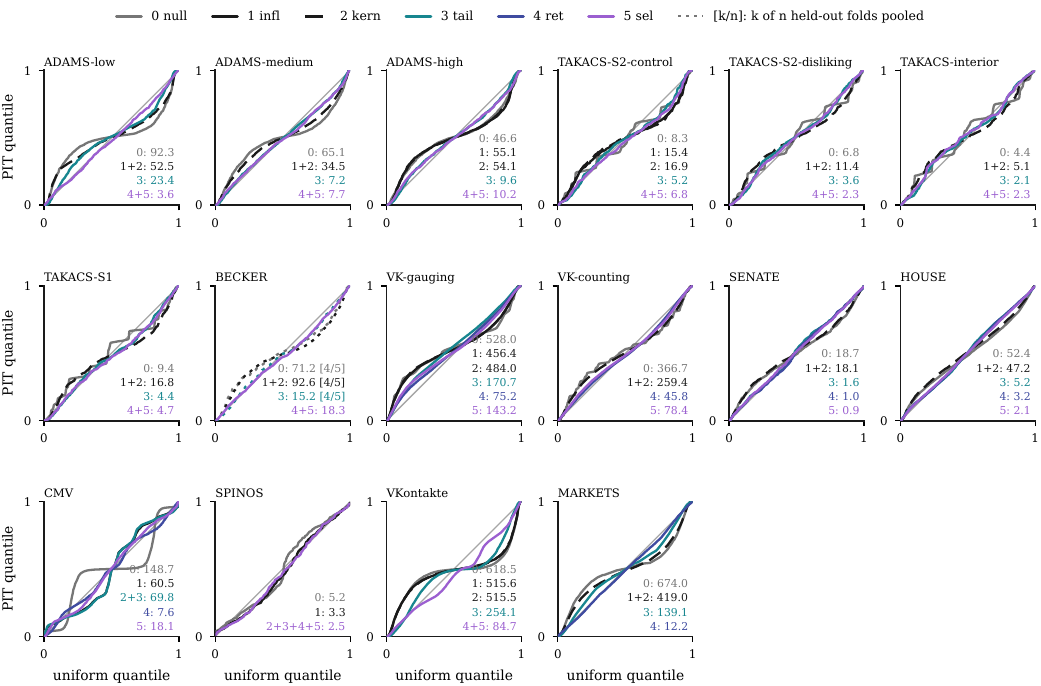}
  \caption{Held-out PIT QQ curves at every rung of the ladder (0, null, to 5, selection) on every dataset. The legend gives each rung's Anderson--Darling statistic; [k/n] marks datasets where only k of n folds could be pooled. The PIT export has no rung 3 for \vkc{} and no rung 5 for \markets{}, so those curves are missing.}
  \label{fig:si_qq_by_rung}
\end{figure}

\begin{figure}[H]
  \centering
  \includegraphics[width=\textwidth,height=0.9\textheight,keepaspectratio]{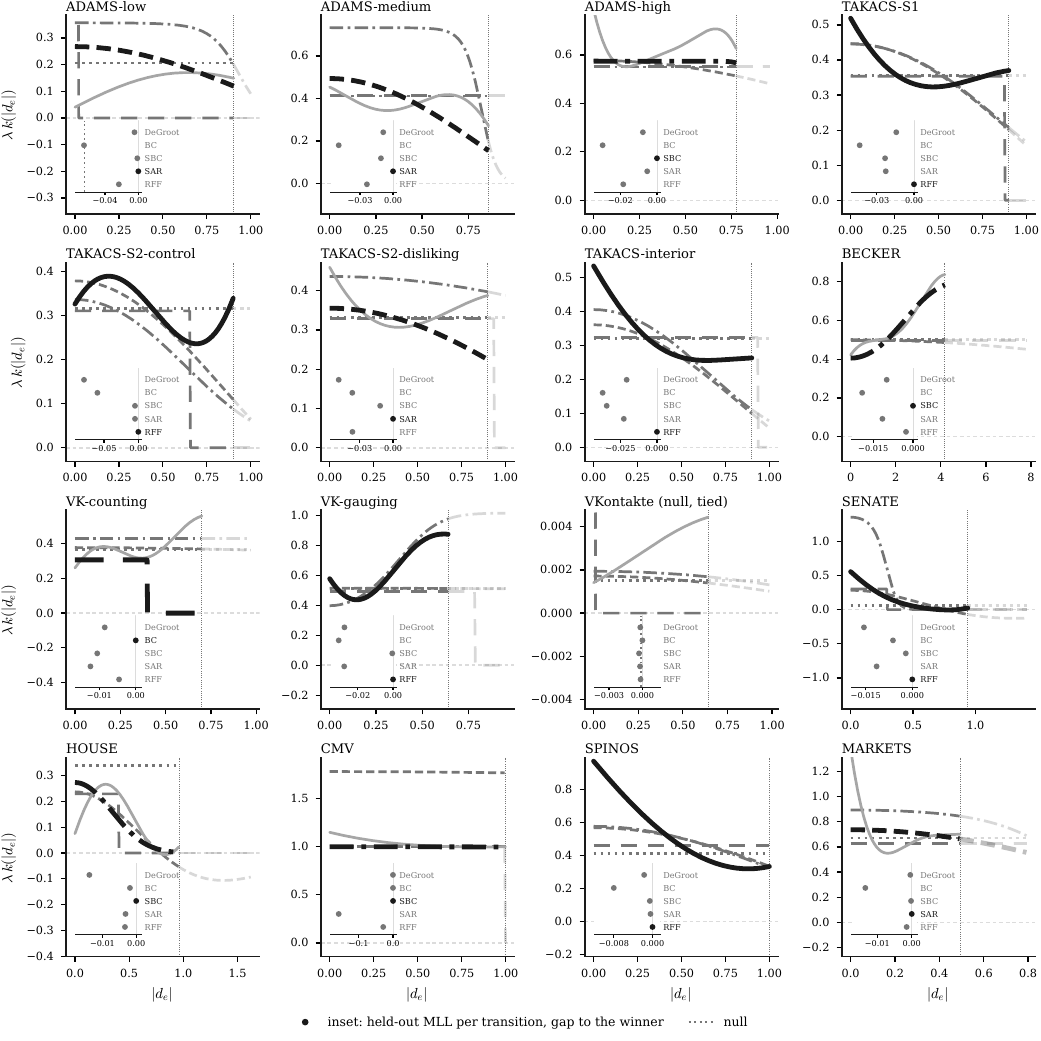}
  \caption{Every kernel's best fit with latent retention on every panel: the dyadic experiments in the first two rows, the networked datasets in the last two. Curves show $\lambda k(|d|)$; the winning kernel is colored and the rest are gray (dashed if parametric). Beyond the 99th percentile of discrepancies (dotted line) the curves are faded. The inset gives each fit's held-out MLL per transition as a gap to the winner, with the null as a dashed line. Where a dataset has both arms, each kernel's fit is the better of the two. On \kozitsin{} the best kernel ties the null (within $10^{-3}$ nats per transition), so no kernel is colored.}
  \label{fig:si_kc}
\end{figure}

\begin{figure}[H]
  \centering
  \includegraphics[width=0.85\textwidth]{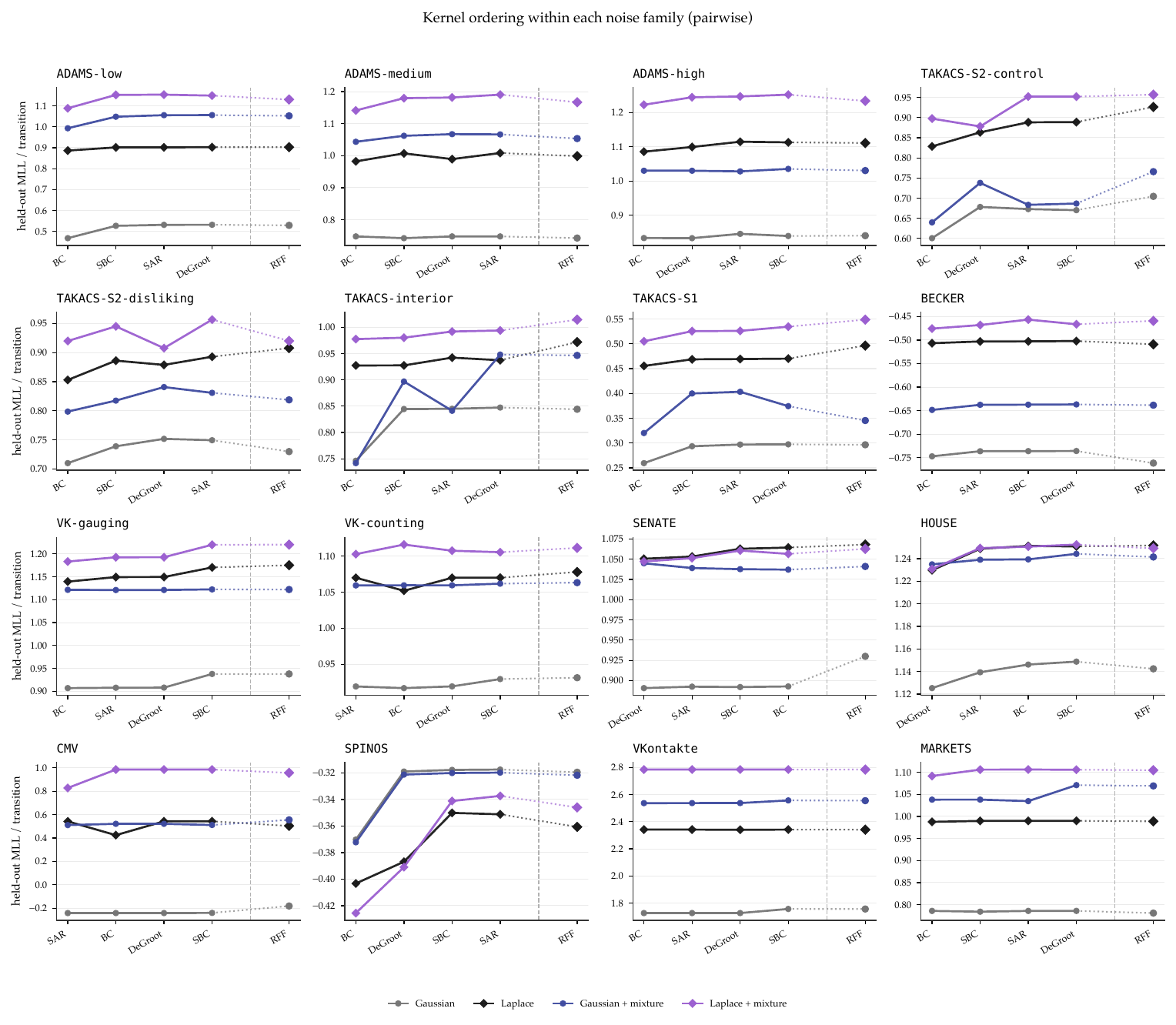}
  \caption{Held-out MLL per transition of every kernel under each of the four noise models, latent-selection arm on the networked datasets. A dyadic experiment has one candidate source, so its two arms are the same model.}
  \label{fig:si_grid_pairwise}
\end{figure}

\begin{figure}[H]
  \centering
  \includegraphics[width=\textwidth]{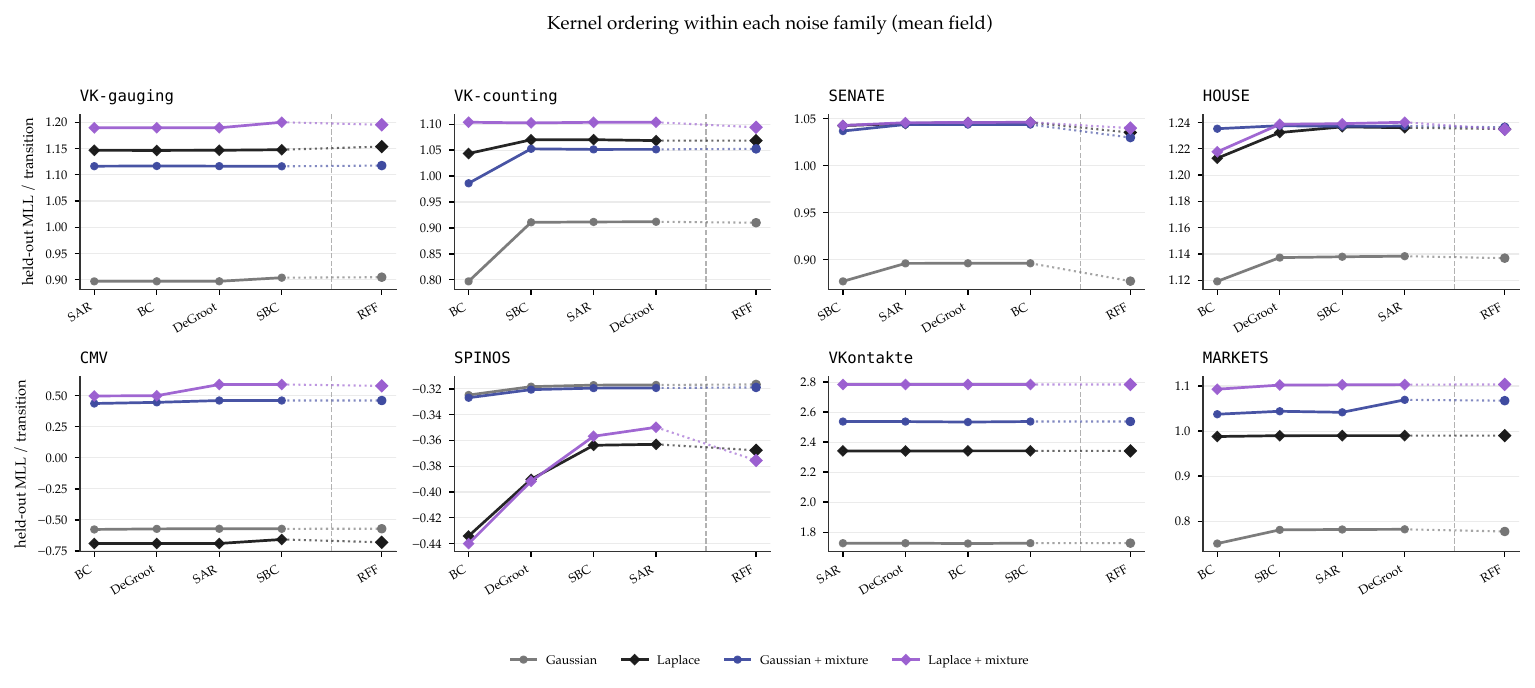}
  \caption{As \Cref{fig:si_grid_pairwise}, for the neighbor-aggregation arm on the networked datasets.}
  \label{fig:si_grid_meanfield}
\end{figure}

While the main text presents selected results, the selected fit for every dataset is reported in \Cref{tab:si_winners_compact} and the full ladder in \Cref{tab:si_ladder_compact}. \Cref{fig:si_qq_by_rung} shows calibration at every rung, \Cref{fig:si_kc} shows every kernel's fit under the full model, and \Cref{fig:si_grid_pairwise,fig:si_grid_meanfield} compare kernels under each noise model.

\begin{table}[H]
  \centering
  {\footnotesize\setlength{\tabcolsep}{2pt}
\begin{tabular}{llrrrrr}
\toprule
dataset & model (kernel, noise, arm) & $\hat\phi$ / $\hat\gamma$ & $\hat\lambda$ & $\hat\eta_1$ & $\hat\pi$ & MLL/transition \\
\midrule
\adamsLo{} & SAR, Laplace mixture & 1.36 & 0.27 & 0.12 & 0.45 & 1.154 \\
\adamsMid{} & SAR, Laplace mixture & 1.13 & 0.49 & 0.077 & 0.58 & 1.191 \\
\adamsHi{} & SBC, Laplace mixture & 0.785, 23.7 & 0.57 & 0.053 & 0.8 & 1.252 \\
\takacsCtl{} & RFF, Laplace mixture & $\hat\gamma=1.25$ & 0.33 & 0.075 & 0.86 & 0.957 \\
\takacsDis{} & SAR, Laplace mixture & 1.72 & 0.36 & 0.076 & 0.9 & 0.957 \\
\takacsInt{} & RFF, Laplace mixture & $\hat\gamma=1.25$ & 0.53 & 0.068 & 0.93 & 1.015 \\
\takacsOne{} & RFF, Laplace mixture & $\hat\gamma=1$ & 0.52 & 0.12 & 0.93 & 0.549 \\
\becker{} & SBC, Laplace mixture & 0.0718, -0.182 & 0.82 & 0.22 & 0.8 & -0.457 \\
\vkg{} & RFF, Laplace mixture, latent selection & $\hat\gamma=2.78$ & 0.58 & 0.049 & 0.75 & 1.220 \\
\vkc{} & BC, Laplace mixture, latent selection & 0.4 & 0.31 & 0.06 & 0.93 & 1.116 \\
\senate{} & RFF, Laplace, latent selection & $\hat\gamma=1.8$ & 0.53 & 0.057 &  & 1.068 \\
\house{} & RFF, Laplace, latent selection & $\hat\gamma=2.29$ & 0.073 & 0.048 &  & 1.252 \\
\cmv{} & SBC, Laplace mixture, latent selection & 1.15, 40.4 & 1 & 0.046 & 0.7 & 0.985 \\
\spinos{} & RFF, Gaussian, neighbor aggregation & $\hat\gamma=0.5$ & 0.98 & 0.39 &  & -0.317 \\
\kozitsin{} & null (tied), Laplace mixture &  &  & 0.027 & 0.63 & 2.784 \\
\markets{} & SAR, Laplace mixture, latent selection & 1.89 & 0.69 & 0.08 & 0.32 & 1.106 \\
\bottomrule
\end{tabular}}

  \caption{Selected (highest held-out MLL) fit for each dataset: kernel, noise model and, for networked datasets, arm. The shape column gives $\hat\phi$: one width for BC and SAR, a width and a slope for SBC, and the bandwidth $\hat\gamma$ for \rff{}. $\hat\lambda$ is the kernel scale. $\hat\eta_1$ is the scale of the influence component: the standard deviation for Gaussian noise and the scale for Laplace noise. $\hat\pi$ is the influence probability among nonzero updates (exact zeros are removed before fitting), blank when the selected noise model has no retention (\senate{}, \house{}, \spinos{}). The last column is the held-out MLL per transition. Every parameter comes from one fit, the CV fold with the highest training MLL, the same fit the kernel figures draw. A dataset whose best kernel beats the random-walk null by less than $10^{-3}$ nats per transition is reported as the null (tied), with no kernel parameters (\kozitsin{}). Likewise a model with retention that beats the best model without it by less than $10^{-3}$ nats is reported without retention (\house{}). This comparison is made with selection in place; at the retention rung of \Cref{tab:si_ladder_compact}, before selection, retention still gains 0.004 nats on \house{}. SBC is fit without a positivity constraint, so its slope can invert the gate (\becker{}).}
  \label{tab:si_winners_compact}
\end{table}

\begin{table}[H]
  \centering
  {\small\setlength{\tabcolsep}{2.5pt}
\begin{tabular}{lrrrrrrlr}
\toprule
panel & infl. & shape & tail & ret. & sel. & total & rung 2 & AD final (rung) \\
\midrule
\adamsLo{} & 0.062 & 0.000 & 0.371 & 0.247 & --- & 0.680 & DeGroot & 3.6 (5) \\
\adamsMid{} & 0.155 & 0.000 & 0.251 & 0.183 & --- & 0.589 & DeGroot & 7.7 (5) \\
\adamsHi{} & 0.298 & 0.007 & 0.271 & 0.134 & --- & 0.710 & RFF & 10.2 (5) \\
\becker{} & 0.362 & 0.000 & 0.234 & 0.043 & --- & 0.639 & DeGroot & 18.3 (5) \\
\cmv{} & 0.276 & 0.001 & 0.000 & 1.151 & 0.405 & 1.834 & RFF & 18.1 (5) \\
\house{} & 0.034 & 0.000 & 0.098 & 0.004 & 0.013 & 0.148 & DeGroot & 2.1 (5) \\
\kozitsin{} & 0.006 & 0.000 & 0.615 & 0.443 & 0.000 & 1.064 & RFF & 84.7 (5) \\
\markets{} & 0.040 & 0.000 & 0.208 & 0.114 & 0.002 & 0.364 & DeGroot & 12.2 (4) \\
\vkc{} & 0.143 & 0.000 & 0.156 & 0.035 & 0.008 & 0.343 & DeGroot & 78.4 (5) \\
\vkg{} & 0.179 & 0.008 & 0.249 & 0.042 & 0.025 & 0.502 & RFF & 143.2 (5) \\
\senate{} & 0.020 & 0.000 & 0.150 & 0.000 & 0.022 & 0.192 & DeGroot & 0.9 (5) \\
\spinos{} & 0.076 & 0.002 & 0.000 & 0.000 & 0.000 & 0.078 & RFF & 2.5 (5) \\
\takacsCtl{} & 0.239 & 0.026 & 0.222 & 0.031 & --- & 0.517 & RFF & 6.8 (5) \\
\takacsDis{} & 0.273 & 0.000 & 0.156 & 0.012 & --- & 0.441 & DeGroot & 2.3 (5) \\
\takacsInt{} & 0.287 & 0.000 & 0.125 & 0.043 & --- & 0.454 & DeGroot & 2.3 (5) \\
\takacsOne{} & 0.201 & 0.000 & 0.199 & 0.052 & --- & 0.452 & DeGroot & 4.7 (5) \\
\bottomrule
\end{tabular}}

  \caption{Gain in held-out MLL per transition at each step of the capability ladder: influence (rung 1), kernel shape (rung 2), heavy tail (rung 3), retention (rung 4) and selection (rung 5). Dyadic datasets have no selection step (---). Total is the gain over the random-walk null. \emph{Rung 2} is the kernel selected at rung 2. \emph{AD final} is the Anderson--Darling statistic of the held-out PIT against the uniform at the highest rung for which it is computed (rung in brackets).}
  \label{tab:si_ladder_compact}
\end{table}

\begin{figure}[H]
  \centering
  \includegraphics[width=0.9\textwidth]{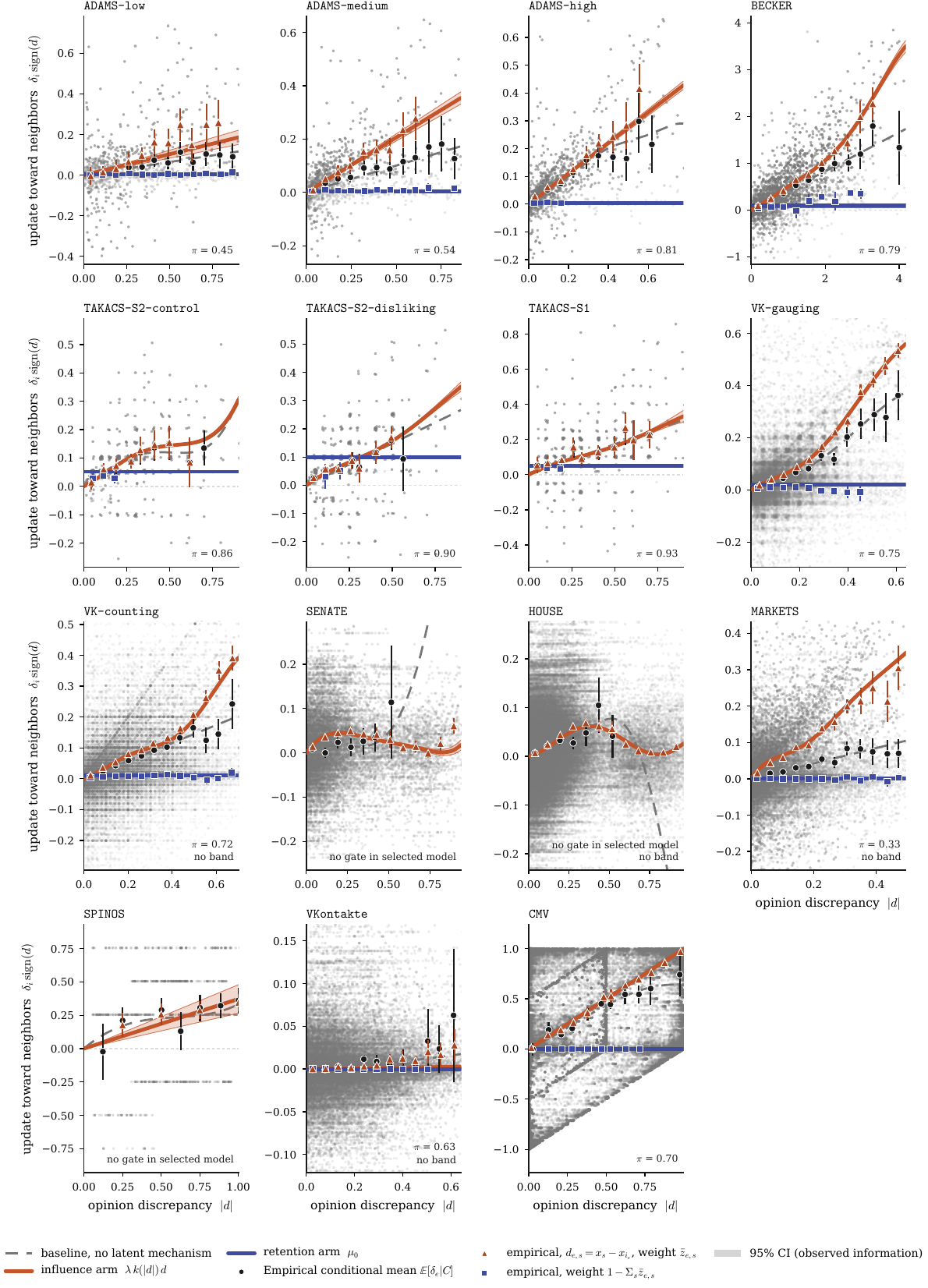}
  \caption{Fitted kernels on every dataset, as in \Cref{fig:kernel_comparison}. The baseline kernel (no latent mechanism, dashed) and the influence arm $\hat\lambda\,\kernel(d)$ of the selected model (orange), with binned empirical means. As in \Cref{fig:kernel_comparison}, the selected model is the better of DeGroot and \rff{}, not the full-family winner of \Cref{tab:si_winners_compact}. Bands are 95\% Fisher intervals on the observed information; \vkc{}, \house{}, \markets{} and \kozitsin{} are drawn without bands.}
  \label{fig:si_kernel_arms_all}
\end{figure}

\clearpage
\subsection{Kernel and noise grids}



\end{document}